\documentclass{aa}
\pdfoutput=1
\usepackage[utf8]{inputenc}
\usepackage{graphicx}

\begin{document}

\title{Near-infrared scattering as a dust diagnostic}

\author{Mika Saajasto\inst{1}
  \and Mika Juvela\inst{1}
  \and Johanna Malinen\inst{1}$^{,}$\inst{2}}


\institute{Department of Physics, P.O.Box 64, FI-00014, University of Helsinki, Finland
\and Institute of Physics I, University of Cologne, Germany
}

\date{Received day month year / Accepted day month year}

\abstract
{Regarding the evolution of dust grains, from diffuse regions of space to dense molecular cloud cores, many questions remain open. Scattering at near-infrared wavelengths, or 'cloudshine', can provide information on cloud structure, dust properties, and radiation field that is complementary to mid-infrared 'coreshine' and observations of dust emission at longer wavelengths.}
{We examine the possibility of using near-infrared scattering to constrain the local radiation field and the dust properties, the scattering and absorption efficiency, the size distribution of the grains, and maximum grain size.}
{We use radiative transfer modelling to examine the constraints provided by J, H, and K bands in combination with mid-infrared surface brightness at 3.6 $\mu$m. We use a spherical one-dimensional and elliptical three-dimensional cloud models to study the observable effects of different grain size distributions with varying absorption and scattering properties. As an example, we analyse observations of a molecular cloud in Taurus, TMC-1N.}
{The observed surface brightness ratios between the bands change when the dust properties are changed. However, even a small change of $\pm 10\%$ in the surface brightness of one channel changes the estimated powerlaw exponent of the size distribution $\gamma$ by up to $\sim 30\%$ and the estimated strength of the radiation field $K_{\rm ISRF}$ by up to $\sim 60\%$. The maximum grain size $A_{\rm max}$ and $\gamma$ are always strongly anti-correlated. For example, overestimating the surface brightness by $10 \%$ changes the estimated radiation field strength by $\sim 20 \%$ and the exponent of the size distribution by $\sim 15\%$. The analysis of our synthetic observations indicates that the relative uncertainty of the parameter distributions are on average $A_{\rm max}, \gamma \sim 25\%$, and the deviation between the estimated and correct values $\Delta Q < 15 \%$. For the TMC-1N observations, a maximum grain size $A_{\rm max} > 1.5$ $\mu$m and a size distribution with $\gamma > 4.0$ have high probability. The mass weighted average grain size is $\langle a_{\rm m} \rangle = 0.113$ $\mu$m, which is $\sim 20\%$ higher compared to the MRN distribution.}
{We show that scattered infrared light can be used to derive meaningful limits for the dust parameters. However, errors in the surface brightness data can result in considerable uncertainties in the derived parameters.}


\maketitle

\section{Introduction}

The exact composition and size distribution of interstellar dust are still an open question. The properties of dust are known to evolve from diffuse interstellar medium (ISM) to dense regions \citep{Bernard1999, Ridderstad2006} It has been assumed that in diffuse regions the properties of dust are uniform. However, the Planck-HFI observations showed significant variations in dust spectral energy distribution (SED), in the long-wavelength opacity spectral index all over the sky \citep{PlanckXXIV2011, PlanckXI2014, PlanckXVII2014, PlanckXXIX2016}. \citet{Ysard2015} showed that the variations of the SED can not be explained by small changes in the radiation field but rather by variation or evolution of dust properties.

Evidence of dust grains growing in size in dense regions of the ISM was reported by \citet{Stepnik}, using dust emission observations of a filament in the Taurus cloud complex covering a wavelength region between 200 - 600 $\mu$m. \citet{Stepnik} concluded that the observations could only be explained by taking into account larger grains \citep{Rawlings2005,Ysard2012,Martin2012,Ysard2013}.

Scattering at near-infrared (NIR) wavelengths can be used to study the composition of interstellar dust. The first observations of surface brightness at NIR wavelengths were reported by \citet{LehtinenMattila1996}. \citet{Foster2006} showed that it is possible to obtain large maps of the surface brightness and named the phenomenon 'cloudshine'. The history of light scattering observations in dark clouds is covered in more detail by \citet{Juvela2006}. The sensitivity to the size of the scattering particles makes observations of scattered light potentially useful in the study of the evolution of grain sizes.

\citet{Coreshine} studied molecular cloud L183 and reported an excess surface brightness in mid-infrared (MIR) towards the dense core. The MIR excess was also discovered in a larger sample of cloud cores by \citet{Pagani}, and \citet{GCCIII}. The surprisingly high surface brightness in the Spitzer 3.6 $\mu$m band was named 'coreshine'. It was interpreted to trace light scattering caused by larger grains than indicated by the classical size distribution presented by \citet{MRN} (hereafter MRN). \citet{Andersen} studied the scattered light of cloud core L260, and came to a conclusion that, in order to explain the shapes of the surface brightness profiles, grain sizes up to 1 $\mu$m were required. Other recent studies on coreshine have considered the location on the sky and the strength of the coreshine effect \citep{Global} and the effect of turbulence as a cause for enhanced grain growth and coreshine \citep{Turbulence}. The detection of coreshine is considered a direct evidence of dust grains growing in size in the ISM, because confirming the growth from thermal emission observations requires detailed modelling, see however, \citet{Ysard2016}.

Coreshine has been detected from numerous sources above and below the galactic plane \citep{Global}, and in some cases near the galactic anti-center. Detection of coreshine towards the bright galactic plane is difficult due to the strong background combined with the extinction caused by the core \citep{Global}. Additionally, the unknown properties of the local radiation field cause problems in reliable confirmation of coreshine. In a recent paper, \citet{Lefevre2016} studied the possibility to use longer wavelengths, up to 8 $\mu$m, to place constraints on dust properties. The authors showed that, the surface brightness of the scattered light at 8 $\mu$m is brighter than has previously been thought, and scattering at those longer wavelengths should also be taken into account.

If the grain sizes are originally below 0.2 $\mu$m, the grain size should affect the NIR colours. In this study we explore the limitations of using NIR and NIR combined with MIR observations to probe the properties of interstellar dust. Apart from Spitzer space telescope and before The James Webb Space Telescope, there are no space borne observatories capable of observing at MIR wavelengths. Since ground-based MIR observation are difficult, the more readily available NIR observations could be a crucial tool in the study of the dense ISM.

Radiative transfer computations can be used to determine how the properties of dust grains and of the local radiation field translate into observed surface brightness. From the modelled surface brightness, we can estimate confidence limits for dust parameters, for example, the maximum grain size and the powerlaw exponent of the size distribution. In this study, we will use radiative transfer calculations to derive surface brightness for model clouds in the J, H, K, and 3.6 $\mu$m bands. The calculations use a dust model with varying size distributions. We will use a one-dimensional spherical model and a three-dimensional ellipsoid cloud model. The results of our radiative transfer computations will be analysed with the Markov chain Monte Carlo (MCMC) method to estimate confidence regions for the parameters of the grain size distribution. As an example of real observations and to compare the simulations to the observations, we will utilize NIR and MIR observations of the filament TMC-1N \citep{Malinen12, Malinen} in the Taurus molecular cloud complex, covering the J, H, K, 3.6$\mu$m, and 4.5 $\mu$m bands.
\citep{Malinen12} studied the possibility of using NIR observations to derive filament profiles and \citep{Malinen} studied the possibility of using scattered NIR light on large scale to study the properties of filamentary structures, compared to, for example, dust emission studies.  

The content of this paper is as follows: In Sect. \ref{Sect:2}, we give an overview of the radiative transfer and MCMC methods and present the cloud models we use in our study. We present our main results in Sect. \ref{Sect:3} and discuss the results in Sect. \ref{Sect:4}. Finally, in Sect. \ref{Sect:5} we summarise our findings. The TMC-1N observations are summarised in Appendix A.

\section{Methods}\label{Sect:2}

We investigate how the observed surface brightness depends on the local radiation field and the dust properties. In the following we describe the cloud models and the radiative transfer methods used in this study.

\subsection{Cloud models}\label{Sect:2.1}

We use spherically symmetric one dimensional cloud model and a three-dimensional ellipsoid model. In one-dimensional spherical cloud model the density distribution is set according to Bonnor-Ebert model ($\xi=4.5$, $M_{\odot}=25$) \citep{Bonnor1956, Sipila2015}. The model optical depth $\tau$ is varied by a direct multiplication of the cloud densities. The model is divided into 50 shells such that the innermost 20 shells contain approximately 40 $\%$ of the mass.

The three-dimensional elliptical cloud model is a prolate ellipsoid with an axis ratio of 3:1, that is discretised onto a Cartesian grid of 68$^3$ cells. The density profile is defined as

\begin{equation}
\rho_i = \exp( -k_1 X^2 - k_2({Y}^2 + {Z}^2))
\end{equation}
where X, Y, and Z are the Cartesian coordinates of the point $i$ and $k$ = $4 \times \ln(2) / L_{\rm ax}^2$, where $L_{\rm ax}$ is the length of the axis.

\subsection{Radiative transfer}\label{Sect:2.2}

We assume that the surface brightness consists of scattered radiation and of background radiation seen through the cloud. In particular, we assume that there is no notable thermal emission. The assumption follows the arguments provided by \citet{Coreshine}; integrating the infrared emission model of \citet{DL07}, we would expect 10 times more emission in the 5.8 $\mu$m band than in the 4.5 $\mu$m band. In observations, including our data on the Taurus filament (see Appendix A), the intensity of the 4.5 $\mu$m band is observed to be lower than the intensity of the 5.8 $\mu$m band and thus the contribution of emission is not significant. The observed surface brightness excess relative to the background sky is 

\begin{equation}\label{eq:sironta_malli}
I(\lambda) = I_{sca}(\lambda) + I_{bg}(\lambda)e^{-\tau} - I_{bg}(\lambda),
\end{equation}
where $I_{sca}(\lambda)$ is the intensity of the scattered light, $I_{bg}(\lambda)e^{-\tau}$ is the intensity of the background radiation coming through an optical depth of $\tau$, and $I_{bg}(\lambda)$ is the intensity of the background seen around the cloud. Thus we compare the surface brightness of the cloud against the background intensity. In this paper, we do not consider the complication of part of the $I_{bg}(\lambda)$ originating in regions between the cloud and the observer.

In our simulations, dust properties are defined by the absorption and scattering efficiencies $Q_{\rm sca}$, $Q_{\rm abs}$, and the asymmetry parameter of the Henyey-Greenstein scattering function $g$ from \citet{WD01}. Furthermore, we assume a simple powerlaw size distribution, $n(a) \propto a^{- \gamma}$, where $a$ is the grain size, with a constant minimum grain size 1.0 nm, and a maximum grain size $A_{\rm max}$ and powerlaw exponent $\gamma$ which can be varied compared to the reference case with $A_{\rm max} = 0.25$ $\mu$m and $\gamma = 3.5$.

\begin{figure}
\resizebox{\hsize}{!}{\includegraphics{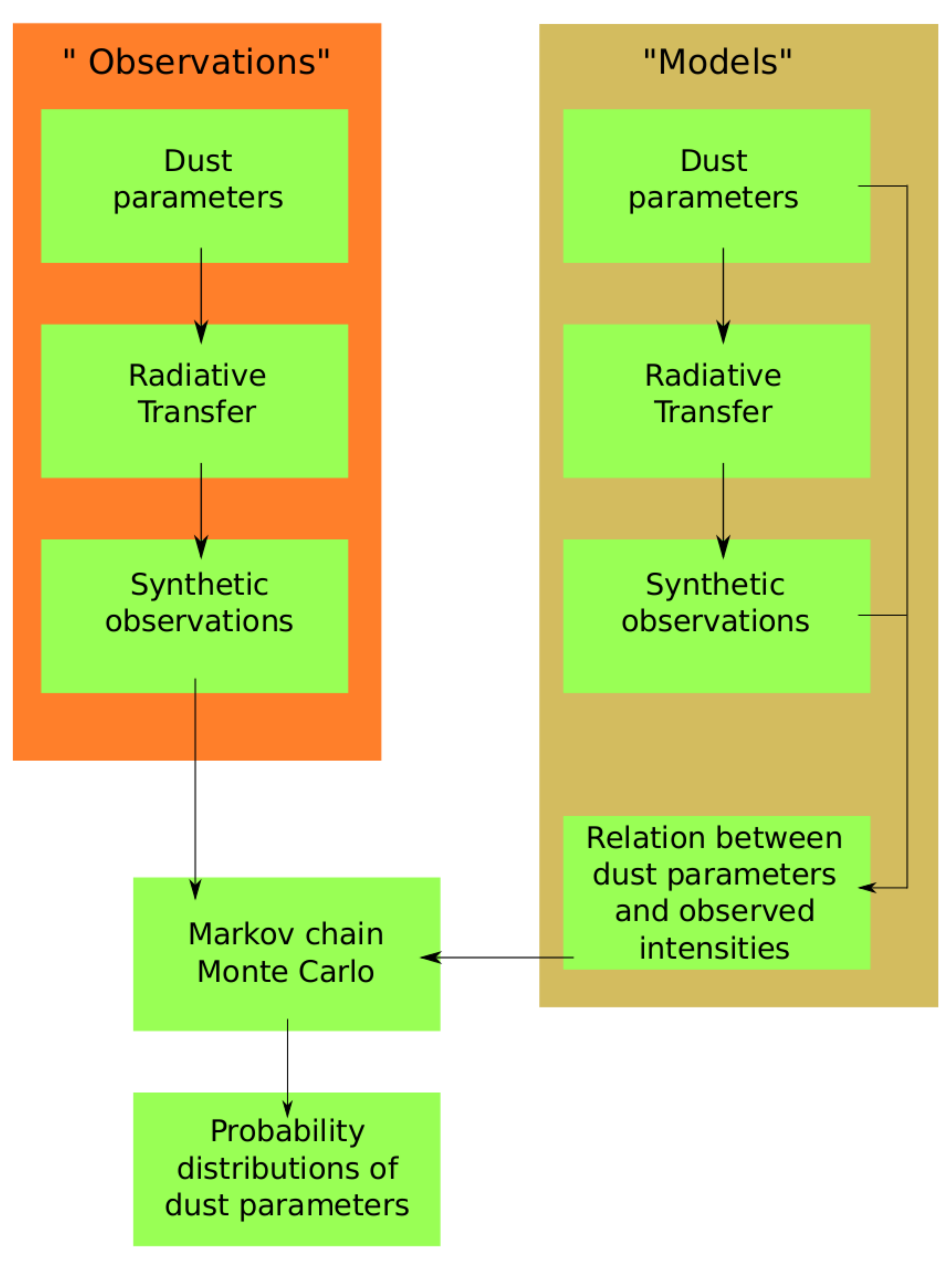}}
\caption{A flow chart describing the steps from dust parameters to probability distributions.}
\label{fig:flow}
\end{figure}

In order to ensure comparability between the models, we normalise the extinction of the J band of both silicate and carbon grains. As a default, we assume that the fraction of the J band optical depth produced by silicate grains is $r_{\rm Si}$=0.5. To examine the sensitivity to differences in the dust properties, we also test cases with $r_{\rm Si}$=0.3 and $r_{\rm Si}$=0.7. Furthermore, because we assume that emission is negligible, we have removed the polycyclic aromatic hydrocarbons (PAHs) from all dust models. To study the effect of increased or decreased scattering efficiency, we scale the albedo, $\alpha$, of the grains by $\pm 10\%$, and in order to study the effect of asymmetry, the value of the asymmetry parameter $g$ is changed by $\pm 10\%$.

We use the radiative transfer code CRT \citep{CRT_1, CRT_2} to compute surface brightness maps for the J, H, K, and 3.6 $\mu$m bands for each combination of $A_{\rm max}$ and $\gamma$, for each band.

\subsection{Parameter estimation}\label{Sect:2.3}

The Markov chain Monte Carlo (MCMC) method is used to draw samples from a probability distribution without having to know explicit shape of the distribution. This is achieved by a random walk. Based on observations (real or synthetic) and model-predicted surface brightness values, we use the MCMC method to sample the probability distributions of the dust parameters.

In the MCMC procedure, one takes a random step in the parameter space and compares the probabilities in the new and old position. If the difference between the probability of the new and old parameter combination is larger than $\ln (R)$, where $R$ is a random number between 0 and 1, the new parameter combination is accepted, otherwise one stays at the old parameters. The current parameter values are saved before repeating the procedure. With enough steps, the saved parameter values trace the probability distributions of the dust parameters.

The radiative transfer computations are performed only for a fixed grid of parameter values, thus, we use linear interpolation to estimate the intensities of scattered light at any value of $A_{\rm max}$ and $\gamma$. The MCMC routine will go trough the probability distribution step-by-step, by taking the interpolated intensity values corresponding to a single value of $A_{\rm max}$ and $\gamma$ at a time. The probability $p$ is

\begin{equation}\label{eq:mcmc}
\ln (p) = -\frac{1}{2} \sum \left( \frac{I_{\rm obs, \textit{i}} -[ I_{\rm sca, \textit{i}}K_{\rm ISRF} - I_{\rm BG, \textit{i}}(1-e^{-\tau_\textit{i}})]}{\sigma_{i}} \right)^2,
\end{equation}
where $I_{\rm obs, \textit{i}}$ are the observed background-subtracted intensities, $I_{\rm BG, \textit{i}}$ are the unattenuated background intensities, $\tau_{i}$ are the optical depths, $\sigma_{i}$ are the assumed observation uncertainties, $I_{\rm sca, \textit{i}}$ are the scattered intensities, and $i$ runs over the bands. Representing the scaling of the ISRF, we use a fourth parameter $K_{\rm ISRF}$ to multiply the $I_{\rm sca, \textit{i}}$ values.

The $\chi^2$ values are

\begin{equation}
\chi^2 = \sum\limits_{i=1}^N (\frac{I_{\rm obs} - I_{\rm mod}}{\sigma_i})^2
\end{equation}
where $I_{\rm obs}$ are the 'observed' (input) intensities, $I_{\rm mod}$ are the corresponding intensities predicted by our model, $\sigma_{\rm i}$ are the uncertainties, and $N$ is the number of bands.

The models are illuminated either by an isotropic radiation field corresponding to a reference interstellar radiation field (ISRF)\citep{ISRF}, or by an anisotropic radiation field derived from the DIRBE all-sky maps (see \citet{Malinen} for details).

We use flat prior distributions, $0.28 - 2.5$ $\mu$m for $A_{\rm max}$, $2.0 - 5.0$ for $\gamma$, and $0.3 - 3.0$ for $K_{\rm ISRF}$. For $\tau_{\rm J}$, we use a value of 6.0 or 2.0 (corresponding to our reference observations, see Sect. \ref{Sect:2.4}). The optical depths of the other bands are determined by the extinction curve of the dust model. The radiative transfer calculations were carried out with a fixed external radiation field, but as the scattered light depends on this linearly, the result can be scaled to any value of ISRF. We use the same $K_{\rm ISRF}$ value for all bands.

\begin{figure}
\resizebox{\hsize}{!}{\includegraphics{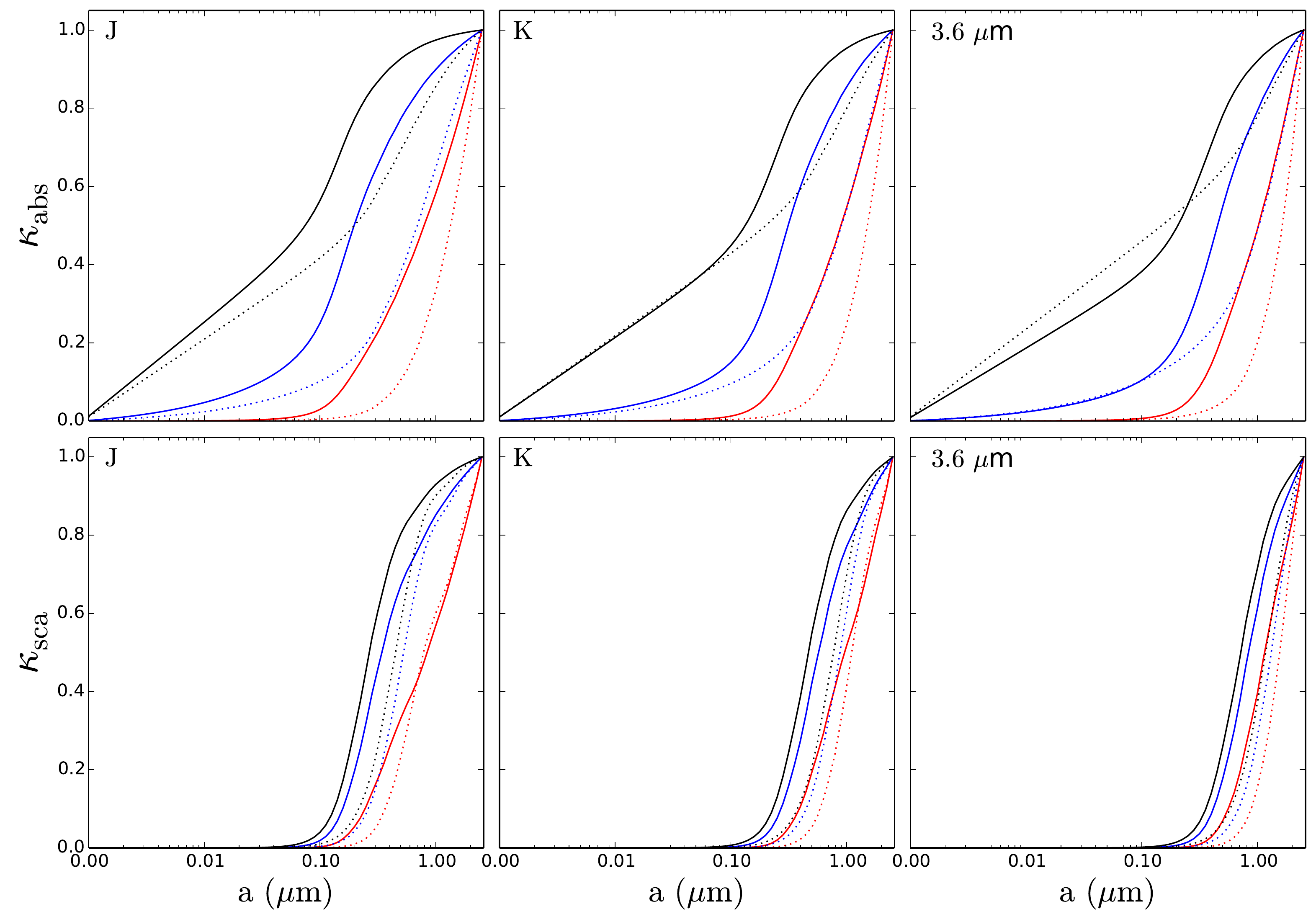}}
\caption{Cumulative sums of absorption and scattering efficiencies for graphite (solid lines) and silicate (dotted lines) grains at J, K, and 3.6 $\mu$m bands. The red lines correspond to $\gamma = 2.5$, the blue lines to $\gamma = 3.5$, and the black lines to $\gamma = 4.5$.}
\label{fig:gra_sil}
\end{figure}

In the fitting, we assume a 10$\%$ uncertainty for the NIR bands. For the 3.6 $\mu$m band the uncertainty is 25$\%$, to take into account the typically higher uncertainties in the background estimation (see appendix A). We also examine a case with 20$\%$ uncertainty for all four bands. We refer to these two cases as $\sigma_{\rm NIR} = 10\%$ and $\sigma_{\rm NIR} = 20\%$, respectively.

\begin{figure*}
\sidecaption
\includegraphics[width=12cm]{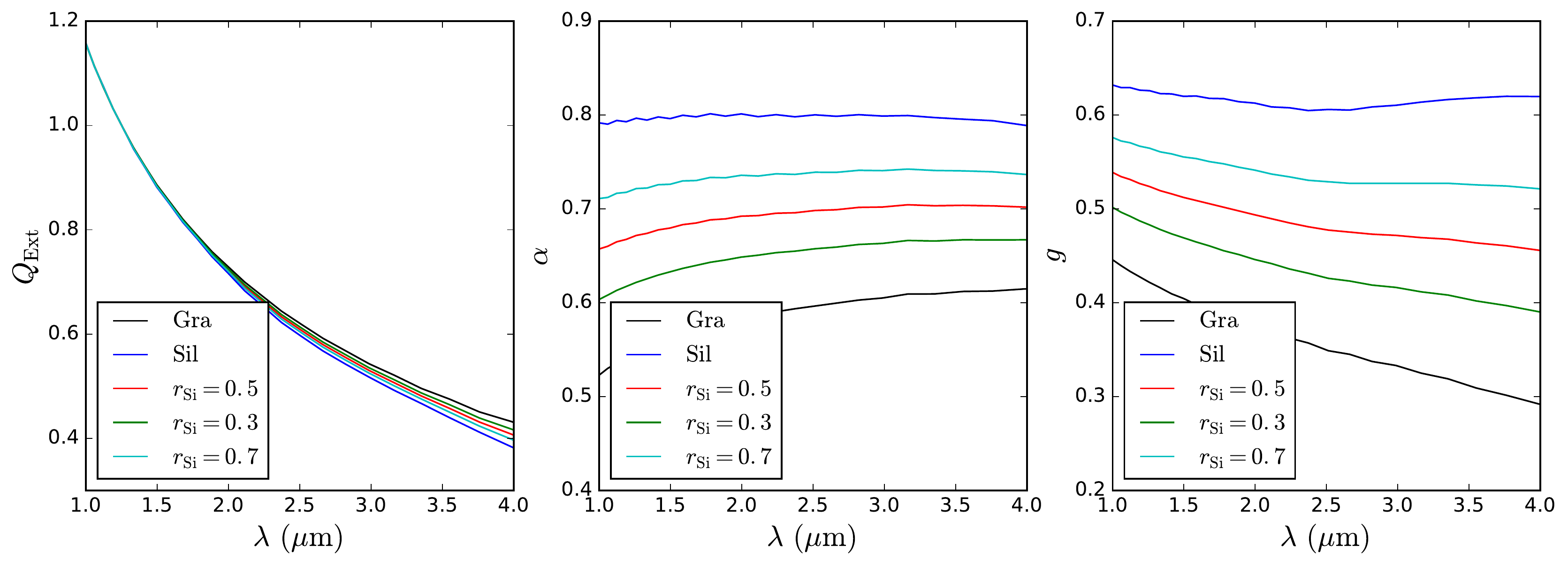} 
\caption{Extinction efficiency, $Q_{\rm ext}$, albedo, $\alpha$, and asymmetry parameter, $g$, for different dust mixtures. The optical properties are normalised so that for J band $Q_{\rm ext} = 1$.}
\label{fig:dustpar}
\end{figure*}

Shown in Fig. \ref{fig:flow} is a summary of the work flow from dust parameters to probability distributions.

\subsection{Taurus observations}\label{Sect:2.4}

\begin{figure*}
\sidecaption
\includegraphics[width=12cm]{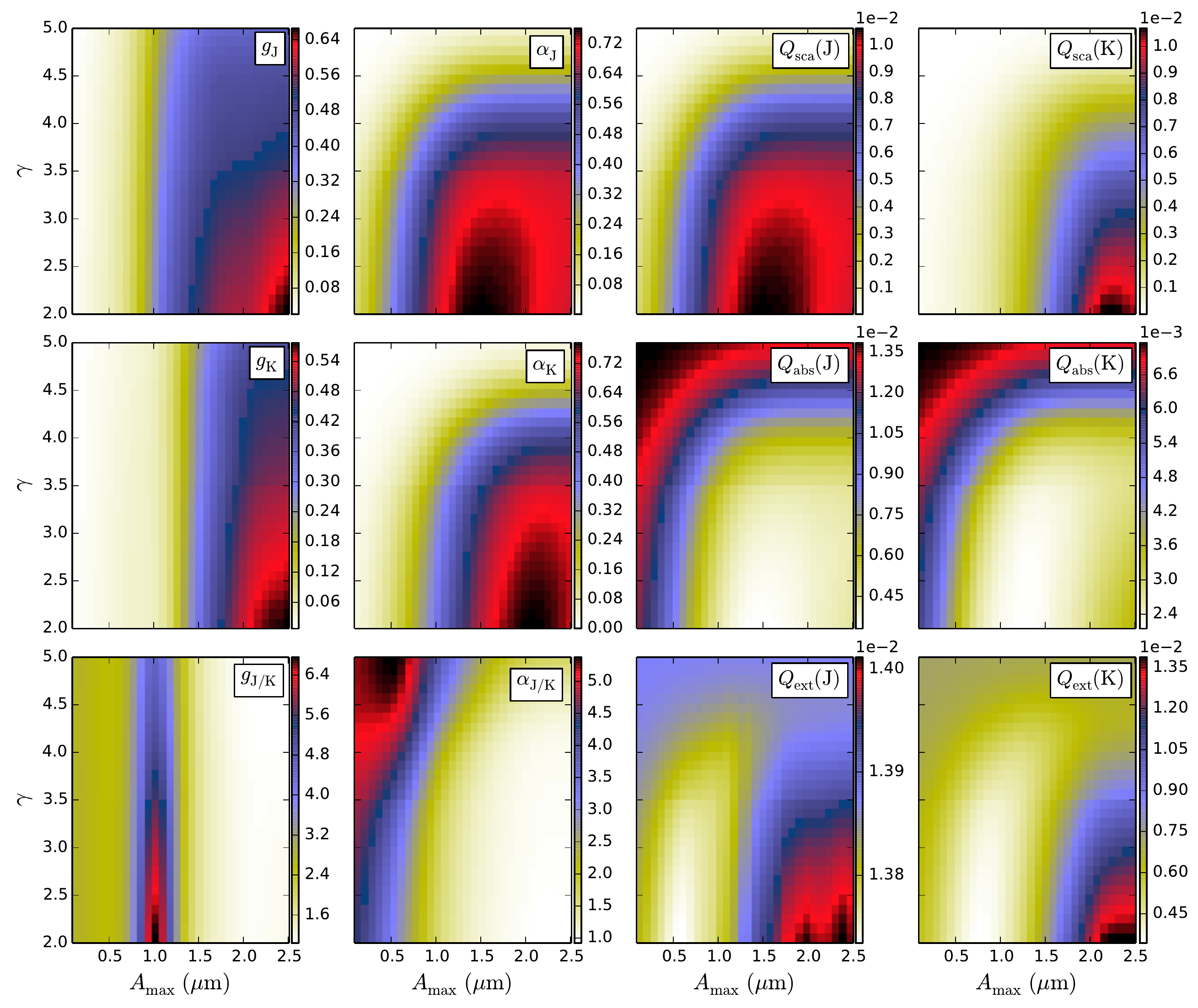} 
\caption{Grain properties as the function of the parameters $A_{\rm max}$ and $\gamma$ of the grain size distribution for the J, and K bands and for the ratio J/K. The panels show the asymmetry parameter, $g$, albedo, $\alpha$, and the scattering, absorption, and extinction efficiencies, $Q_{\rm sca}$, $Q_{\rm abs}$, $Q_{\rm ext}$, respectively. The plot corresponds to $r_{\rm Si}$=0.5 case and the colourbars show the values of the dust parameters.}
\label{fig:opticals}
\end{figure*}

Our study is based mainly on simulated data. However, as an example of real observations, we analyse surface brightness measurements of a filament in Taurus, TMC-1N. The filament was examined by \citet{Malinen12,Malinen}, using the Wide Field CAMera (WFCAM) instrument of the United Kingdom InfraRed Telescope (UKIRT). The observations consist of data in the J, H, and K bands. In addition, we use data from the Spitzer InfraRed Array Camera (IRAC) \citep{Spitzer}. A short overview of the observation is provided in appendix A. For a more detailed description of the observations and data reduction see \citet{Malinen12,Malinen}.

The two chosen positions A and B of the Taurus filament are shown in Fig. \ref{fig:vertailu}. The surface brightness and optical depth values are listed in Table \ref{tab:observations}.

\section{Results}\label{Sect:3}

Before looking at synthetic observations in Sect. 3.2, we examine how the dust cross sections and asymmetry parameters depend on the parameters $\gamma$ and $A_{\rm max}$ of the grain size distribution. 

\subsection{Basic effects of grain size distribution}\label{Sect:3.1}

Shown in Fig. \ref{fig:gra_sil} are the cumulative sums of the absorption and scattering efficiencies of graphite and silicate grains as a function of grain size. For each grain size the absorption and scattering efficiency is defined as

\begin{equation}
\kappa = \pi a^{2} n(a) Q(\lambda),
\end{equation}
where $a$ is the grain size, $n(a)$ is the number density of grains of size $a$, and $Q(\lambda)$ is the absorption or scattering efficiency of the grains of size $a$ at wavelength $\lambda$. Light scattering from grains smaller than $\sim 0.1$ $\mu$m is insignificant, however, the absorption caused by small grains ($a <0.01 \mu$m) can reach $\sim 25 \%$ with large values of $\gamma$.

The extinction efficiency, albedo, and asymmetry parameter for pure silicate and graphite grains and for our three  models with $r_{\rm Si}$=0.5, $r_{\rm Si}$=0.3, $r_{\rm Si}$=0.7 are shown in Fig. \ref{fig:dustpar}. The $Q_{\rm ext}$ is almost constant at NIR wavelengths, however, the albedoes and asymmetry parameters vary between $\sim$0.5-0.8 and $\sim$0.3-0.6, respectively.

In Fig. \ref{fig:opticals}, we show the values of $g$, $Q_{\rm abs}$, and $Q_{\rm sca}$ and the resulting albedo, $\alpha$, and extinction efficiency, $Q_{\rm ext}$, for different combinations of $A_{\rm max}$ and $\gamma$, and assume $r_{\rm Si}$=0.5. As expected, the extinction decreases and the scattering efficiency relative to absorption increases, when moving from J to K band. Furthermore, the maxima of the scattering efficiency $Q_{\rm sca}$ shift from $A_{\rm max} \sim 1.5$ $\mu$m to $A_{\rm max} \sim 2.25$ $\mu$m. For both J and K band, the scattering efficiency decreases with increasing $\gamma$ values.

The differences of the asymmetry parameters indicate that the relative brightness between the J and K bands depends strongly on the location of the cloud with respect to the source of illumination. For example, if an optically thin cloud is seen towards the galactic centre, the J band should be bright relative to the K band, if the cloud is seen towards the anti-centre, the K band should be brighter. The effect is caused by the smaller value of the asymmetry parameter of the K band, resulting in stronger back scattering for all dust species with sizes smaller than $\sim$ 1.2 $\mu$m. However, in dense clouds the optical depth of the cloud will have a stronger effect on the relative brightness between the bands. Higher optical depth reduces both the amount of scattered light (relative to dust mass) and the difference between the attenuated and unattenuated background radiation.

\subsection{Synthetic observations}\label{Sect:3.2}

In this section we present the results of synthetic observations produced by radiative transfer calculations with one dimensional models, assuming $r_{\rm Si}$=0.5 and using the reference radiation field described in Sect. \ref{Sect:2.3}.

\begin{figure}
\resizebox{\hsize}{!}{\includegraphics{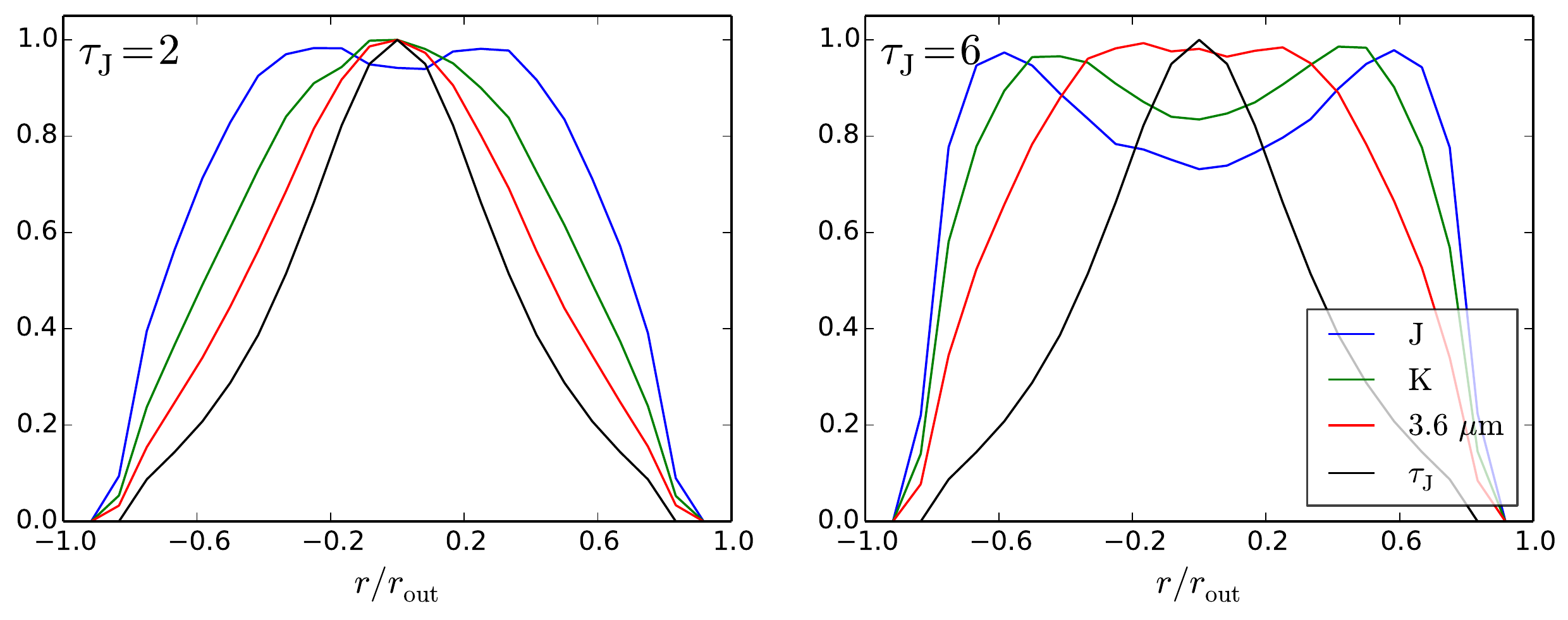}}
\caption{Results for one-dimensional radiative transfer models with $r_{\rm Si}$=0.5, $A_{\rm max} = 1.0$ $\mu$m, and $\gamma = 3.5$ with average $\tau_J = 2$ (left) and $\tau_J = 6$ (right). The panels show the optical depth of the J band and the surface brightness of the J, K, and 3.6 $\mu$m bands. The surface brightness values and optical depth have been normed.}
\label{fig:CRT_center}
\end{figure}

Figure \ref{fig:CRT_center} shows the surface brightness and optical depth profiles computed trough the centre of the model cloud for a dust model with $A_{\rm max}=1$ $\mu$m and $\gamma = 3.5$. The density of the cloud is set so that the average optical depth of the J band computed over the central $r / 10$ region is $\tau_{\rm J} = 2.0$ or 6.0, where $r$ is the radius of the model. The Monte Carlo noise in the computed surface brightness and optical depth values is less than $\sim$1$\%$, significantly less than the typical uncertainties in observations.

\begin{figure*}
\sidecaption
\includegraphics[width=12cm]{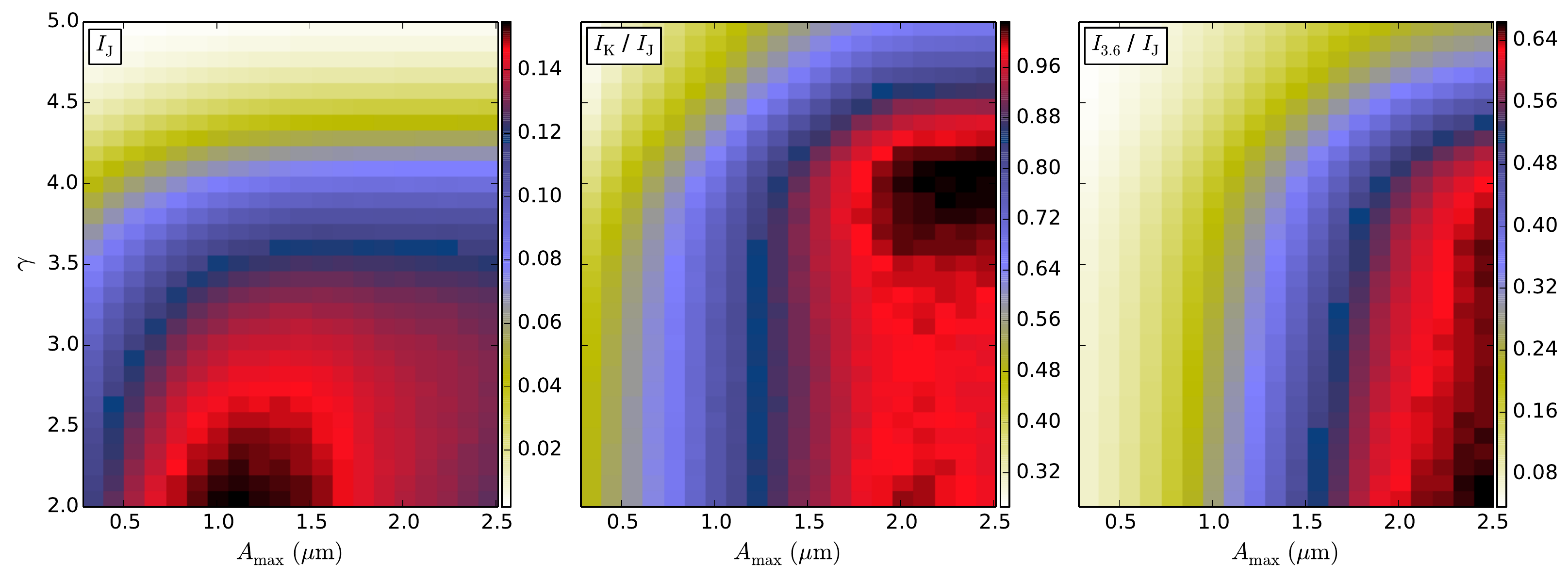} 
\caption{Left: Intensity of the J band as a function of $\gamma$ and $A_{\rm max}$. Centre and right: the intensity ratios between K and J bands, and 3.6$\mu$m and J bands, respectively. A one-dimensional model is used with $r_{\rm Si}$=0.5 and $\tau_{\rm J} = 6$. The colourbars show the intensity and intensity ratios.}
\label{fig:new_intensity}
\end{figure*}

For the $\tau_J=6$ case, the J band surface brightness at the center of the cloud is $\sim 30 \%$ lower than on the outer edge, because of the high optical depth and the resulting saturation of the surface brightness. The K band intensity is $\sim 20\%$ lower in the center. The 3.6 $\mu$m band is starting to saturate and traces more closely the density profile of the cloud. For the $\tau_J=2$ case, only the J band is showing saturation. In Fig. \ref{fig:new_intensity}, we show the intensity of the J band (left), and the intensity ratios between K and J band (centre) and, 3.6 $\mu$m and J band (right) as a function of $\gamma$ and $A_{\rm max}$. As in Fig. \ref{fig:opticals}, because of the small number of large grains, values of $\gamma > 4.5$ produce significantly less intensity, regardless of the value of $A_{\rm max}$. Given that the J band optical depth is fixed, the J band intensity reaches maximum at $A_{\rm max}$ $\sim$ 1.1 $\mu$m and $\gamma \sim 2.22$. Compared to MRN, the distribution contains more large grains as the slope of the size distribution is not as steep.

\begin{figure}
\resizebox{\hsize}{!}{\includegraphics{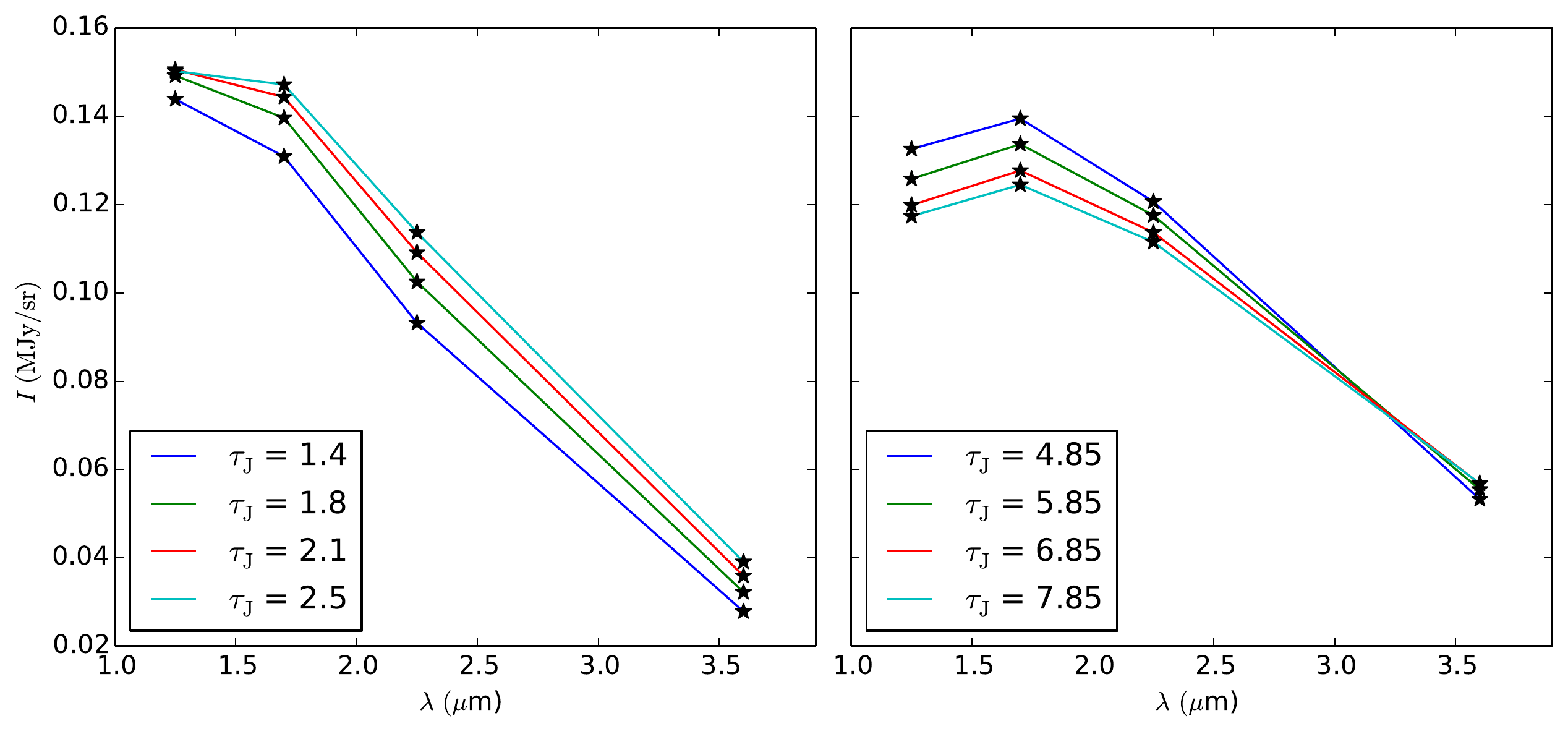}} 
\caption{The spectral energy distributions resulting from a computation with $r_{\rm Si}$=0.5. The maximum grain size is set to $A_{\rm max} = 1.0$ $\mu$m and $\gamma = 3.5$.}
\label{fig:CRT_int}
\end{figure}

The spectral energy distributions with average $\tau_J = 2$ and $\tau_J = 6$ are shown in Fig. \ref{fig:CRT_int}, calculated with $A_{\rm max} = 1.0$ $\mu$m and $\gamma=3.5$. The optical depth of the J band is varied between $1.4 - 2.4$ and between $4.85-7.85$, respectively. The J band begins to saturate when the optical-depth approaches $\sim 2$, as can be seen from the panel on the left. With higher optical depth, right panel, the intensity of J, H, and K channels decreases when the optical depth increases, but the intensity of the 3.6 $\mu$m channel increases.

\begin{figure}
\resizebox{\hsize}{!}{\includegraphics[width=17cm]{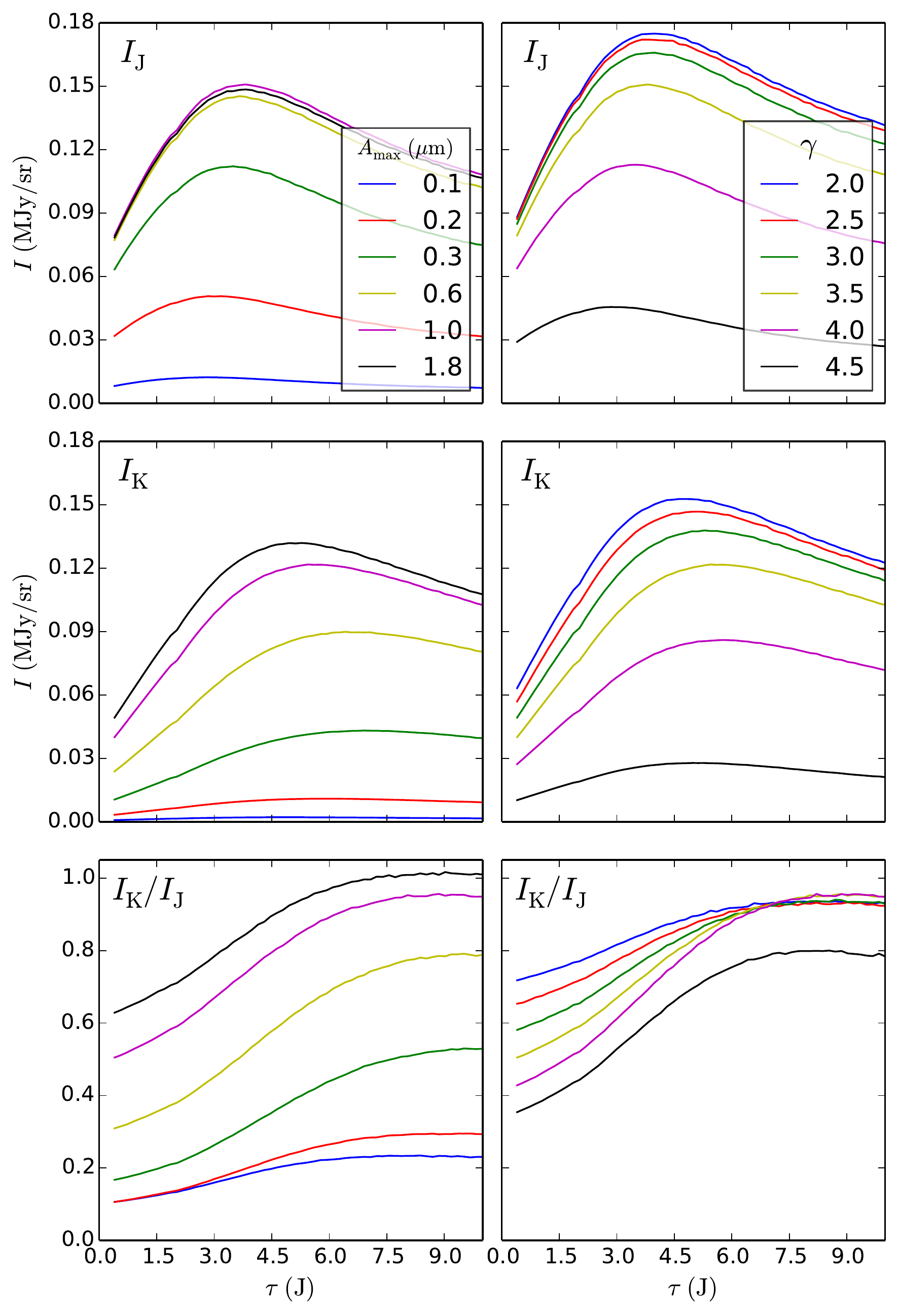}}
\caption{The surface brightness of the J and K bands, and the ratio between K and J bands as a function of optical depth. The colours of the curves correspond to different values of $A_{\rm max}$ and $\gamma$. For the first column, $\gamma = 3.5$, and for the second column $A_{\rm max} = 1.0$ $\mu$m.}
\label{fig:CRT_kayrat}
\end{figure}

The surface brightnesses of the J and K bands and the ratio between J and K bands as a function of optical depth of the J band are shown in Fig. \ref{fig:CRT_kayrat}. The curves correspond to different $A_{\rm max}$ and $\gamma$ values. We show curves for a fixed values of $\gamma = 3.5$, $A_{\rm max}$ varying from 0.1 to 1.8 $\mu$m, and for a fixed value of $A_{\rm max} = 1.0$ $\mu$m, $\gamma$ varying from 2.0 to 4.5.

When $A_{\rm max}$ is constant, the peak intensity is reached at $\tau_{\rm J} \sim 3.8$, for the J band, and $\tau_{\rm J} \sim 5.0$, for the K band, regardless of the value of $\gamma$. Comparing the ratios between the K and J band intensities, it is possible to discern the value of $\gamma$ only if $\tau_{\rm J} < 6$ or $\gamma > 4$.

With a constant $\gamma$, the peak intensity of the J band shifts towards higher optical depth when $A_{\rm max}$ increases, whereas the peak intensity of K band shifts towards lower optical depth when $A_{\rm max}$ increases. It is possible to discern the value of $A_{\rm max}$ from the ratio of K and J band intensities in almost all cases, except if $A_{\rm max} < 0.2$ and the optical depth is low, $\tau_{\rm J} < 2$.

\subsection{MCMC on synthetic observations}\label{Sect:3.3}

To determine the accuracy to which dust parameters can be deduced from observed intensities, we apply the MCMC procedure to synthetic observations of one dimensional and three dimensional clouds where the true dust and ISRF parameters are precisely known. In rough correspondence to the Taurus observations discussed in Sect. \ref{Sect:2.4}, the one dimensional synthetic observations are created using randomly generated values values $\tau_{\rm J}$ = 6.04, $K_{\rm ISRF}$ = 1.53, $A_{\rm max}$ = $1.11 \mu$m and $\gamma$ = 3.49.

\begin{figure}
\resizebox{\hsize}{!}{\includegraphics[width=17cm]{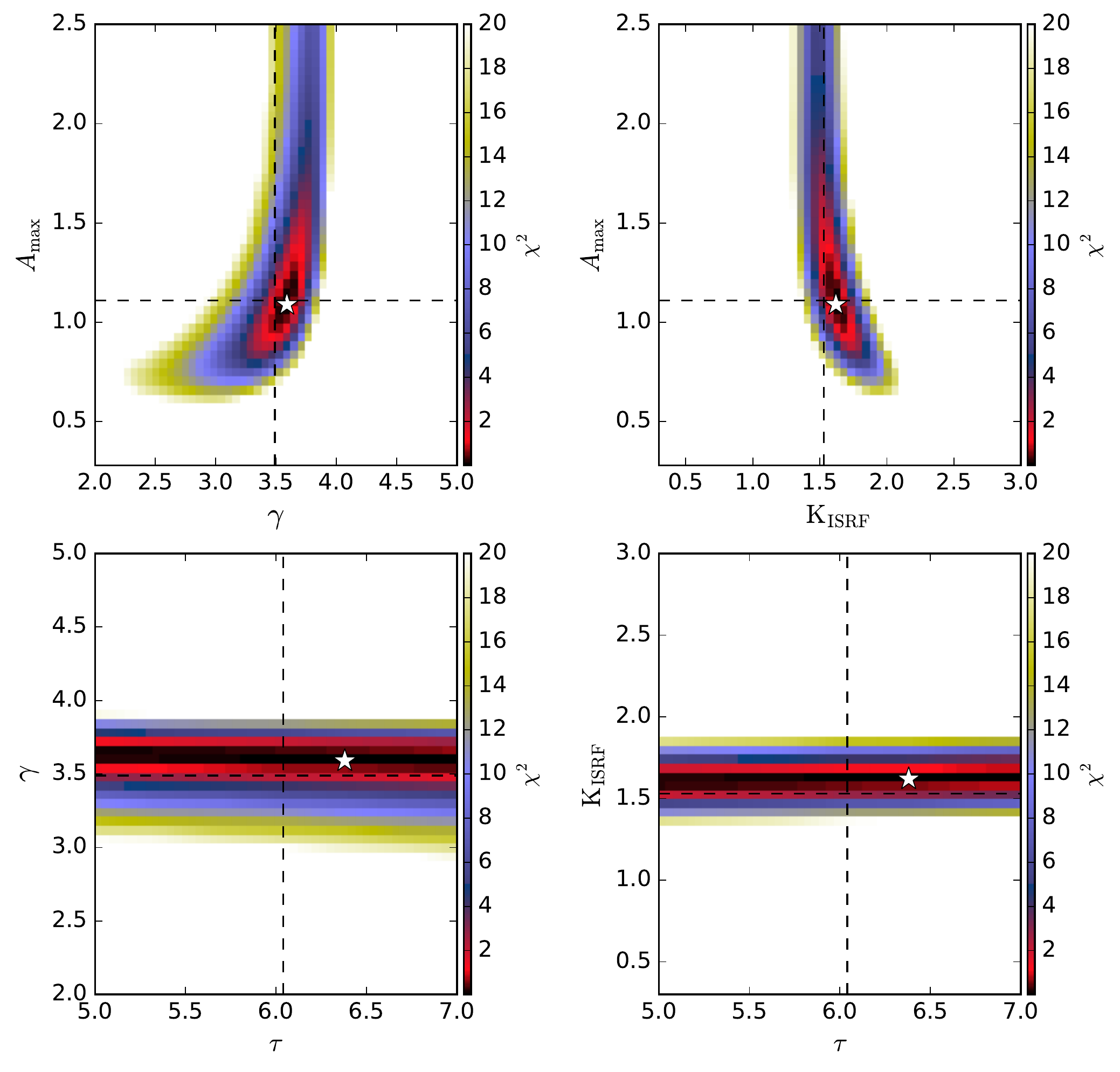}} 
\caption{Cross sections of the $\chi^2$ distribution through the location of the $\chi^2$ minimum for the synthetic observations. The white star indicates the $\chi^2$ minimum and the dashed lines show the dust parameter values used to simulate the synthetic observations. The colourbars correspond to the  $\chi^2$ distribution.}
\label{fig:chimin}
\end{figure}

In Fig. \ref{fig:chimin}, we show the cross sections of the $\chi^2$ distribution using the synthetic observations. The white star marks the location of the minimum at $A_{\rm max} = 1.09, \gamma = 3.53, K_{\rm ISRF} = 1.58$, and $ \tau = 6.36$. The small difference to the input parameters is caused by the use of discretised parameter grids and the shallow $\chi^2$ valley running almost parallel to the $\tau_{\rm J}$ axis. Each parameter grid has 50 points, for $A_{\rm max}$ from 0.25 to 2.5 $\mu$m, for $\gamma$ from 2.5 to 5.0, for $K_{\rm ISRF}$ from 0.3 to 3.0, and for $\tau$ from 4.8 to 7.4. 

\begin{figure*}
\sidecaption
\includegraphics[width=12cm]{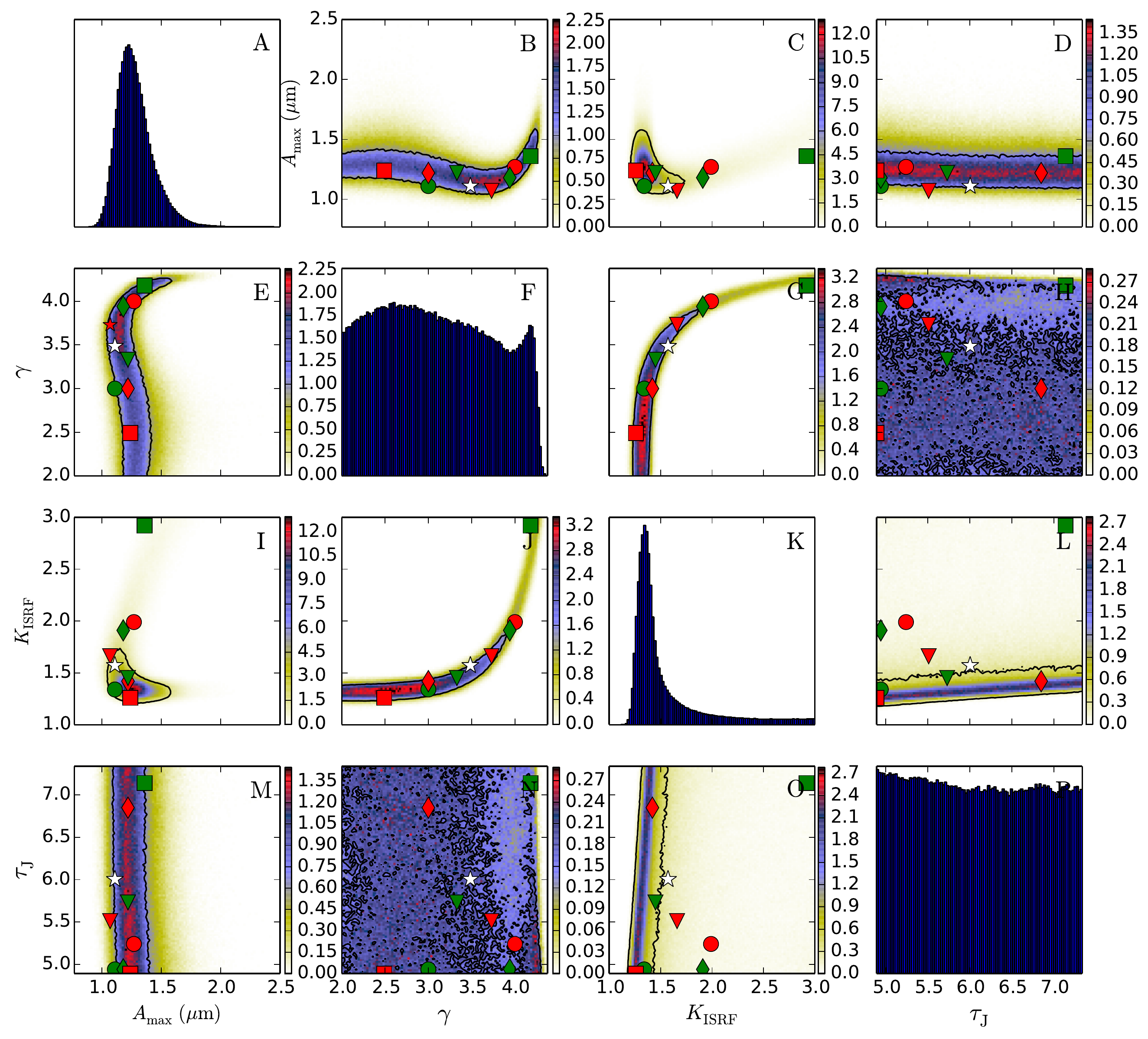} 
\caption{Marginalised one-dimensional probability distributions (panels A, F, K, and P) and two-dimensional projections of the probability of dust parameters for the $\tau_J = 6$ case. The white star indicates the projected position of the $\chi^2$ minimum. The other symbols are the $\chi^2$ minima assuming a 10$\%$ higher or lower intensity for one of the bands, marked with green and red symbols, respectively. The circles correspond to J band, the diamonds to H band, the squares to K band, and the triangles to 3.6 $\mu$m band. The black contour shows the 1 $\sigma$ of the projection. The colour scale shows the normalised probability.}
\label{fig:center_simu_10}
\end{figure*}

Figure \ref{fig:center_simu_10} shows the marginalised probability distributions of a one dimensional model cloud with $\tau_J = 6$. An isotropic radiation field is used in the scattering computations. The distributions are normalised and we only compare the relative probabilities of the different parameter combinations. Figure \ref{fig:center_simu_10} indicates a strong dependence between $\gamma$ and $K_{\rm ISRF}$. The dependence between parameters is also seen in the ($\gamma$, $\tau$) projection as a strong cut-off at high $\gamma$. Thus, to be able to determine either the $\gamma$ or $K_{\rm ISRF}$, the values of the other parameters must already be known with good precision. The four-dimensional parameter space is not easily represented with two-dimensional correlation plots. The maximum of the marginalised probability and the projected location of the minimum in the four-dimensional $\chi^2$ are both shown.

The assumed error estimates of the intensity values have little effect on the results, apart from the broadening of the distributions. Small variations in the intensities of individual channels do affect the location of the $\chi^2$ minimum as can be seen from the locations of the green and red symbols in Fig. \ref{fig:center_simu_10}. Even a moderate change of $\pm$ 10 $\%$ in the intensity of one channel can change the recovered $\chi^2$ minimum values of $\gamma$ and $K_{\rm ISRF}$ by up to $\sim$60\%. In the fit a change of optical depth can be compensated by a change of the radiation field intensity. For further computations we use a constant value of $\tau_{\rm J} = 6$, or $\tau_{\rm J} = 2$, since it is possible to derive a reasonable estimate for the optical depth using other methods, for example with observations of background stars \citep{Nicer}.

Using an anisotropic radiation field in the scattering computations (Fig. \ref{fig:center_3D_isrf_10}), produces well constrained distributions for both $\gamma$ and $K_{\rm ISRF}$ parameters, but the $A_{\rm max}$ distribution is considerably broader and shifted to higher parameter values.

\begin{figure*}
\sidecaption
\includegraphics[width=12cm]{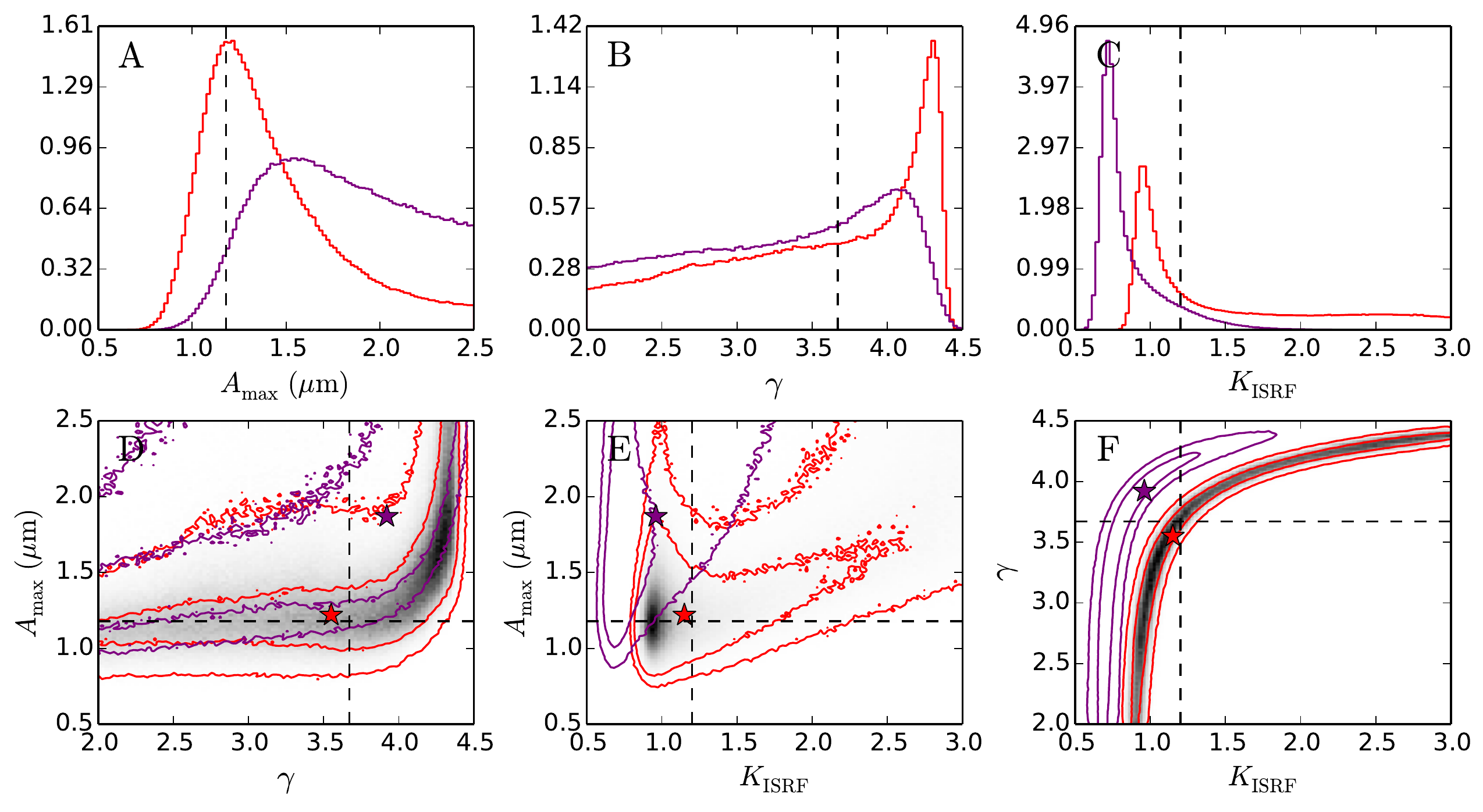} 
\caption{Marginalised probability distributions of dust parameters for an ellipsoidal cloud model. In the radiative transfer modelling (see text), the cloud is viewed either along its minor axis (red lines) or its major axis (purple lines). For both cases, the optical depth along the line-of-sight is $\tau_{\rm J} = 6$ and the dashed lines show the input values used in the computations. The stars indicate the projected positions of the $\chi^2$ minima and the black contours show the 1 and 2 $\sigma$ of the projections. The grey scale map corresponds to the red lines.}
\label{fig:simu_control}
\end{figure*}

The analysis of surface brightness data depends on not only the line-of-sight optical depth but also on the cloud morphology. We derive synthetic observations from a three dimensional prolate ellipsoidal model cloud (see Sect. \ref{Sect:2.1}). The dust parameters were set to $A_{\rm max} = 1.18$, $\gamma = 3.67$, and $K_{\rm ISRF} = 1.2$ and an anisotropic radiation field corresponding to the Taurus case (see \citet{Malinen}) is used in the scattering computations. The optical depth along the line of sight is set to $\tau_{\rm J} = 6.0$.

\begin{figure}
\resizebox{\hsize}{!}{\includegraphics{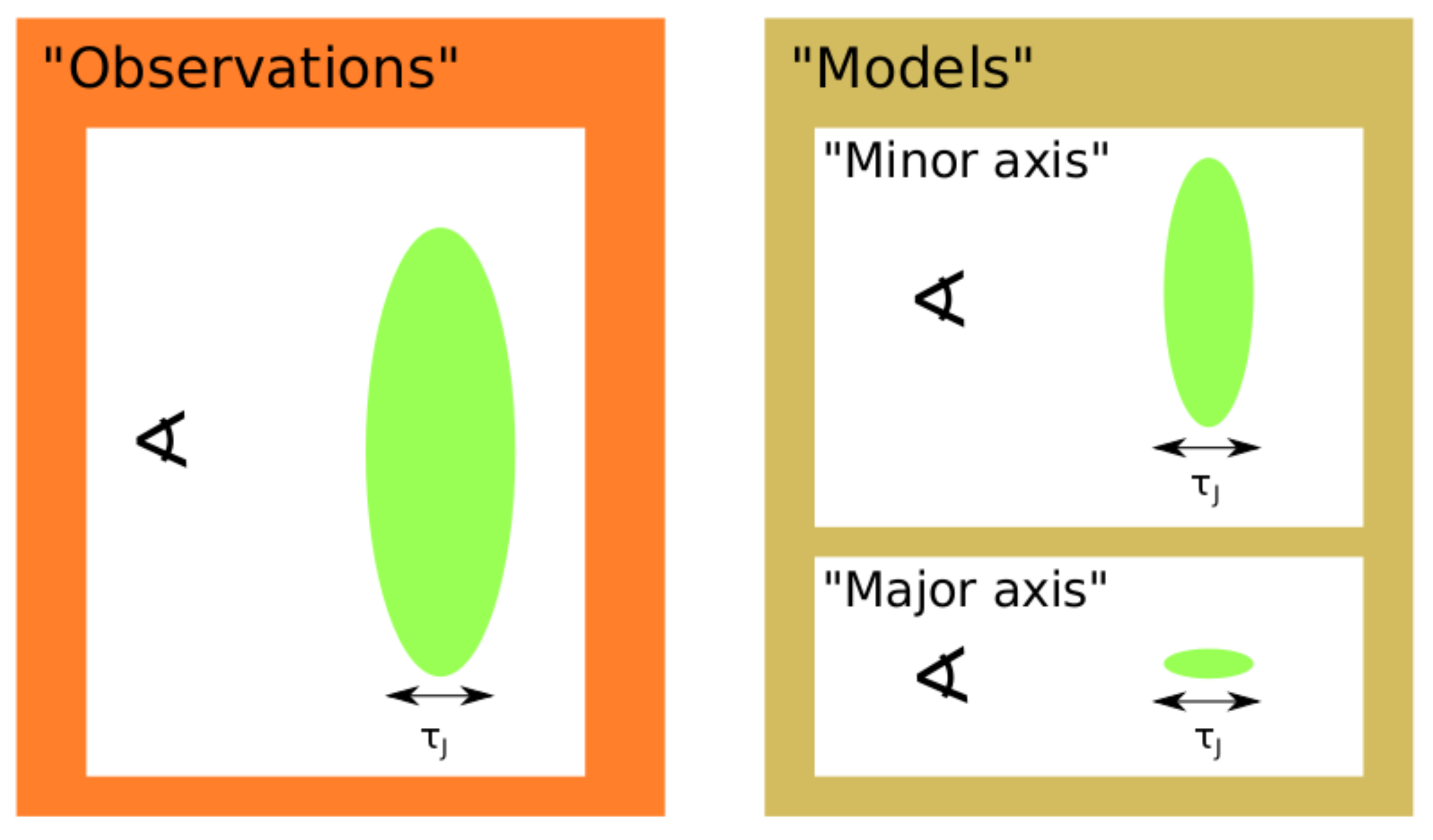}}
\caption{On the right, a schematic of the major axis and minor axis radiative transfer models used to analyse the synthetic observations on the left.}
\label{fig:models}
\end{figure}

For the analysis of these synthetic observations, one needs the relation between dust parameters and the surface brightness. This can be derived with radiative transfer modelling but it further depends on assumptions of the cloud shape. We compare two cases where the relationship is derived with radiative transfer modelling where one views a three dimensional prolate ellipsoidal cloud either along its major axis or its minor axis, see Fig. \ref{fig:models}. In the following we refer to these as the major axis and minor axis radiative transfer models, respectively. In these radiative transfer calculations the model cloud has an axis ratio of 3 and the line-of-sight optical depth is in both cases set to $\tau_{\rm J}=6$. When the radiative transfer model is used to calculate surface brightness along its minor axis, it suffers from more extinction in the direction perpendicular to the line-of-sight (where the full optical depth through the cloud is $\tau_{\rm J=18}$). This affects the radiation field inside the radiative transfer model and thus also leads to a different mapping between the dust parameters and the predicted surface brightness. This is expected to bias the estimates derived for the dust parameters. In the scattering computations we use the anisotropic radiation field. The synthetic observations correspond to the case where the radiative transfer model is viewed along its minor axis. The results with $\sigma_{\rm NIR} = 10\%$ are shown in Fig. \ref{fig:simu_control}, which shows that the assumptions of the radiative transfer models have a clear effect on the estimated dust parameters.

The one-dimensional probability distributions (the minor axis case) of both $\gamma$ and $K_{\rm ISRF}$ are not concentrated around the parameter values that were used to derive the synthetic observations. The maximum of the two-dimensional probability distribution of $\gamma$ and $K_{\rm ISRF}$ are centred around $\gamma = 4.1$, $K_{\rm ISRF} = 1.0$. However, the maxima of the one dimensional $A_{\rm max}$ distribution and the $\chi^2$ minimum are well matched with the input parameter values but the mode of the marginalised probability distributions significantly differs from the location of the $\chi^2$ minimum. In the major axis case, the probability of larger grains is increased significantly as the $A_{\rm max}$ distribution is shifted to $A_{\rm max} > 1.5 \mu$m and the peak of the $\gamma$ distribution is shifted from $\gamma \sim 4.3$ to $\gamma \sim 4.1$. On the other hand, the strength of the radiation field has decreased by $\sim 30\%$.

Increasing the uncertainty of the surface brightness observations, $\sigma_{\rm NIR} = 20\%$, only broadens the resulting probability distributions. Similarly, changes of the asymmetry parameter $g$ do not produce noticeable changes compared to Fig. \ref{fig:simu_control}. 

\begin{figure}
\resizebox{\hsize}{!}{\includegraphics[width=17cm]{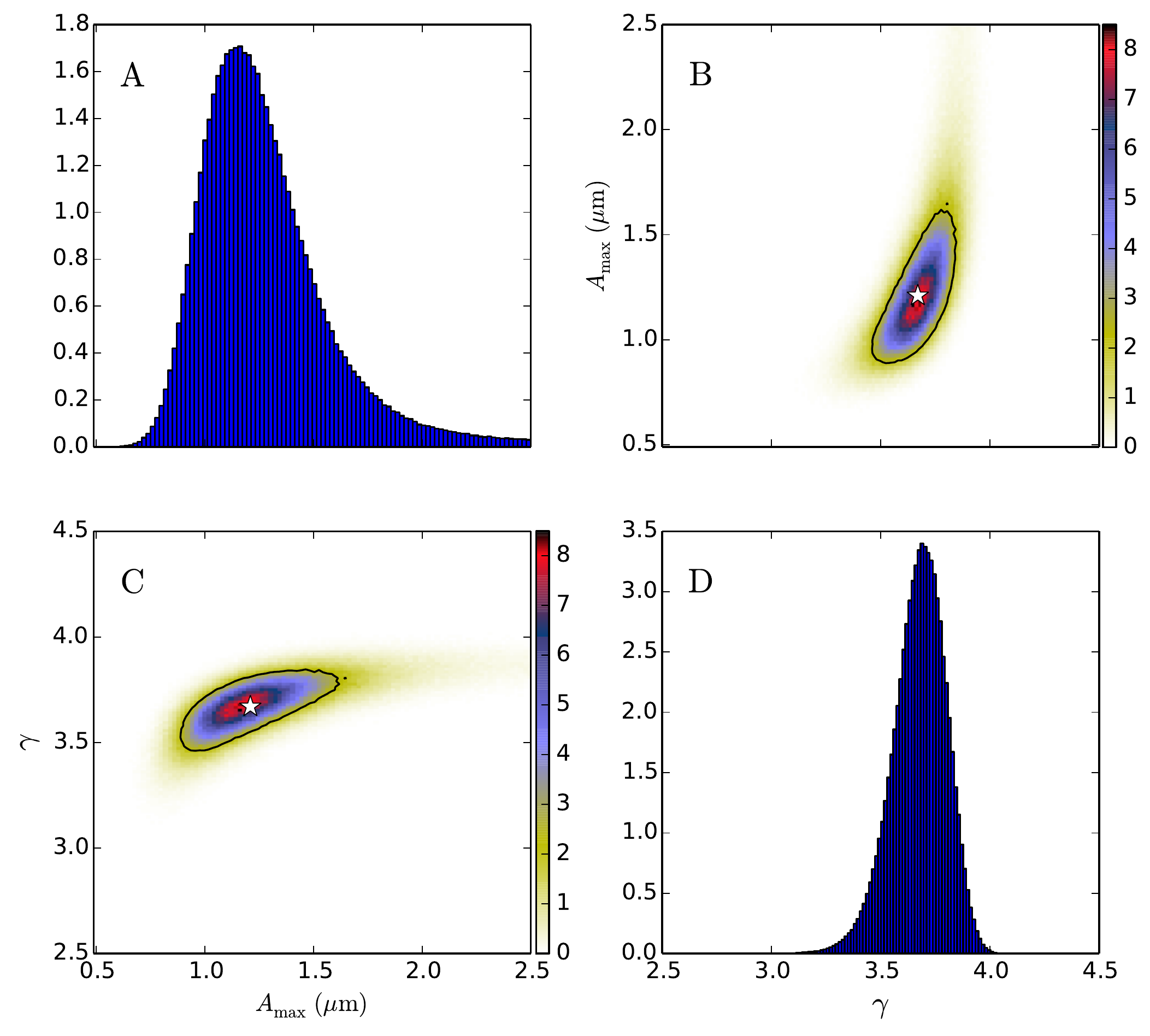}} 
\caption{As Fig. \ref{fig:simu_control}, but the $K_{\rm ISRF}$ is set to a constant value of 1.2.}
\label{fig:no_isrf}
\end{figure}

If $K_{\rm ISRF}$ is fixed to a constant value (Fig. \ref{fig:no_isrf}), the marginalised probability distributions and the $\chi^2$ minimum match the used parameter values with much higher precision.

\begin{figure*}
\sidecaption
\includegraphics[width=12cm]{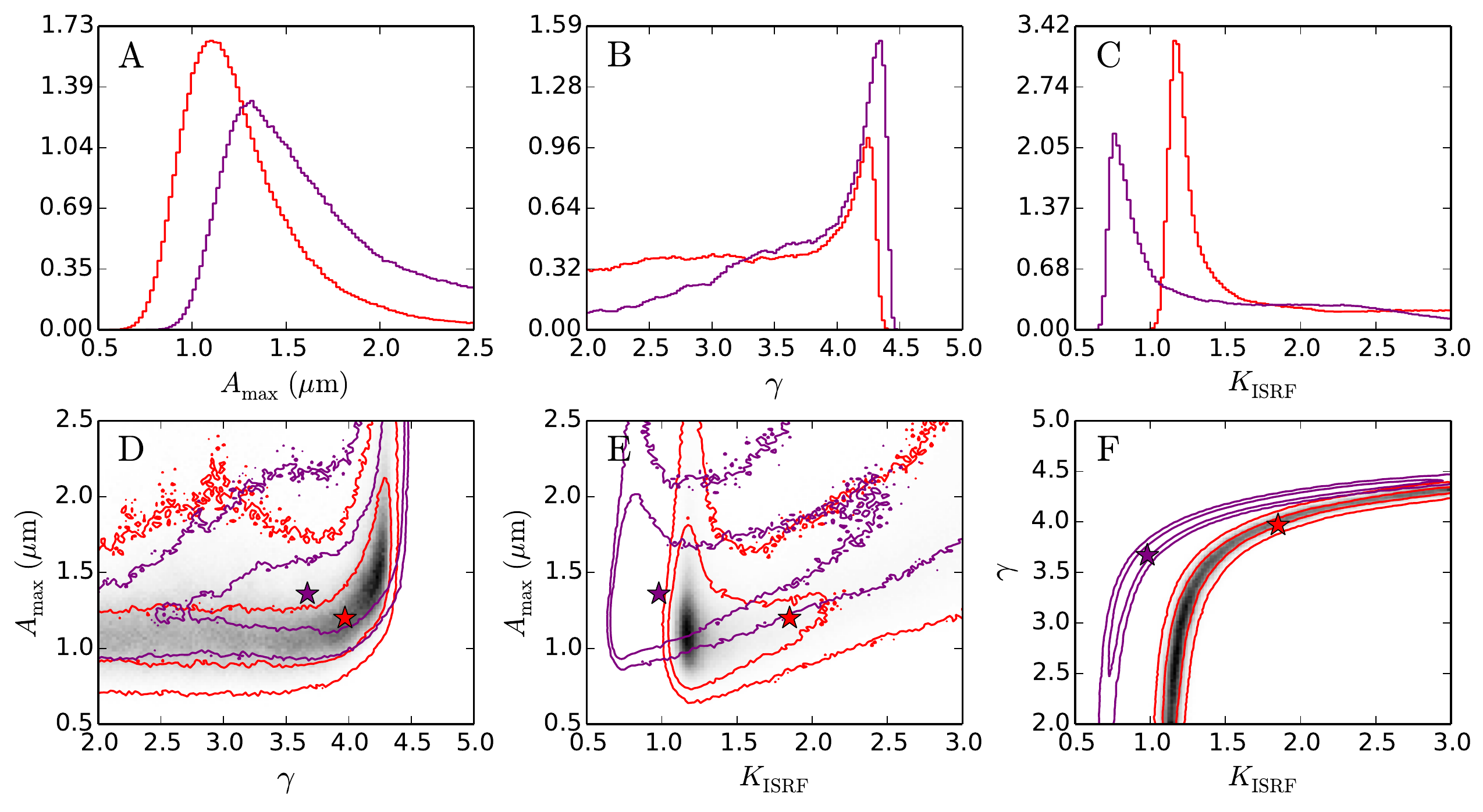}
\caption{Marginalised probability distributions of dust parameters in case $\tau_{\rm J} = 6$. As Fig. 8 but with the grain albedo in the radiative transfer calculations changed by -10\% (red lines) or +10\% (purple lines). The stars indicate the projected positions of the $\chi^2$ minima and the grey-scale map corresponds to the red lines. The dashed lines show the input values used in the computations.}
\label{fig:simu_albedo_mp}
\end{figure*}

Shown in Fig. \ref{fig:simu_albedo_mp} are the results when the albedo is changed by $\pm$10\%, in the scattering simulations, however, for the parameter estimation we use the unmodified albedo. Changes in the albedo have a clear effect to the parameter distributions. Increasing the albedo of the grains shifts the peak of the $A_{\rm max}$ distribution from $A_{\rm max} \sim 1.0$ $\mu$m to $A_{\rm max} \sim 1.5$ $\mu$m, and decreases the probability of $\gamma < 3.5$. Thus, the maximum grain size is increased but the relative amount of large grains is decreased. On the other hand, with higher albedo, the strength of the required radiation field decreases from $K_{\rm ISRF} \sim 1.3$ to $K_{\rm ISRF} \sim 0.7$.

\subsection{The Taurus TMC-1N observations}\label{Sect:3.4}
In this section, we analyse the NIR observations of the Taurus filament described in Appendix A. We use the minor axis cloud model described in Sect. 3.3. For the optical depth of the J band we use values of $\tau_J =  6.0$ and $\tau_J =  2.0$. The surface brightness and optical depth values derived for both cases are shown in Table \ref{tab:observations}. Unless otherwise specified, we use the $r_{\rm Si}$=0.5 ratio.

\subsubsection{MCMC results, $\tau_J =  6$ position}

\begin{figure*}
\sidecaption
\includegraphics[width=12cm]{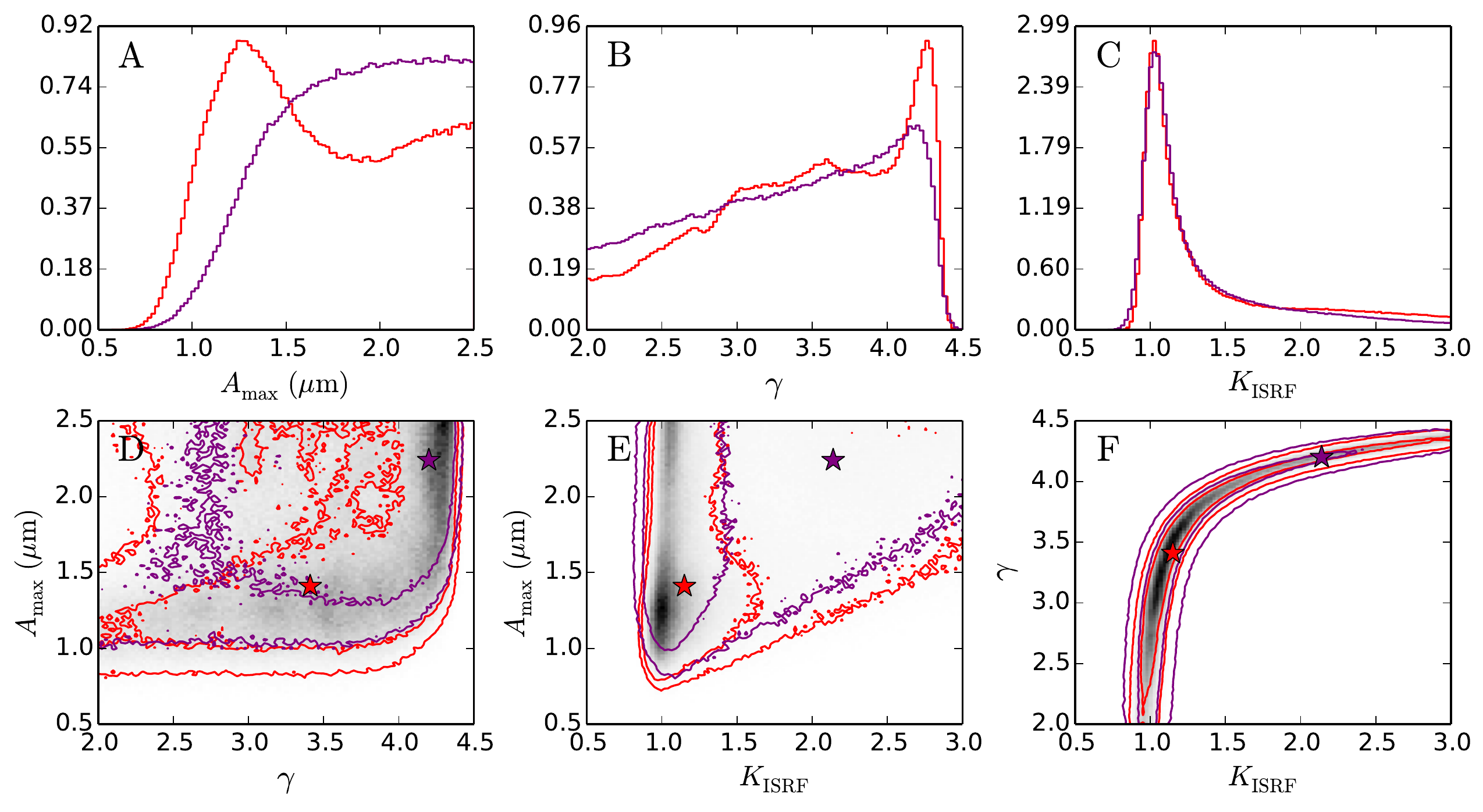} 
\caption{Marginalised probability distributions of dust parameters for the TMC-1N $\tau_J =  6$ position. For the surface brightness data we assume $\sigma_{\rm NIR} = 10\%$, red lines, and $\sigma_{\rm NIR} = 20\%$, purple lines. The stars indicate the projected positions of the $\chi^2$ minima. The grey-scale map corresponds to the red lines and the contours show the 1 and 2 $\sigma$ for both models.}
\label{fig:centre_control_10_20}
\end{figure*}

The results of the MCMC simulations assuming $\sigma_{\rm NIR}=10\%$ or $\sigma_{\rm NIR}=20\%$ where only the $A_{\rm max}$ and $\gamma$ were varied during the scattering computations are shown in Fig. \ref{fig:centre_control_10_20}. A clear peak can be seen at $A_{\rm max}$ $\sim 1 \mu$m and a broad plateau around $\gamma \sim 3.5$. Increase in the $\sigma_{\rm NIR}$ produces smoother parameter distributions. The $\sigma_{\rm NIR}$ has no effect on the $K_{\rm ISRF}$ distribution, however, with higher $\sigma_{\rm NIR}$ $A_{\rm max}$ distribution is pushed to higher values, $ A_{\rm max} \ga 1.5 \mu$m. 

The location of the $\chi^2$ minimum matches reasonably with the maximum of the marginalised probability distributions. However, the minimum seems to follow mainly the $\gamma$ distribution as the minimum is not well matched in the projections with $K_{\rm ISRF}$ or $A_{\rm max}$.

\begin{figure*}
\sidecaption
\includegraphics[width=12cm]{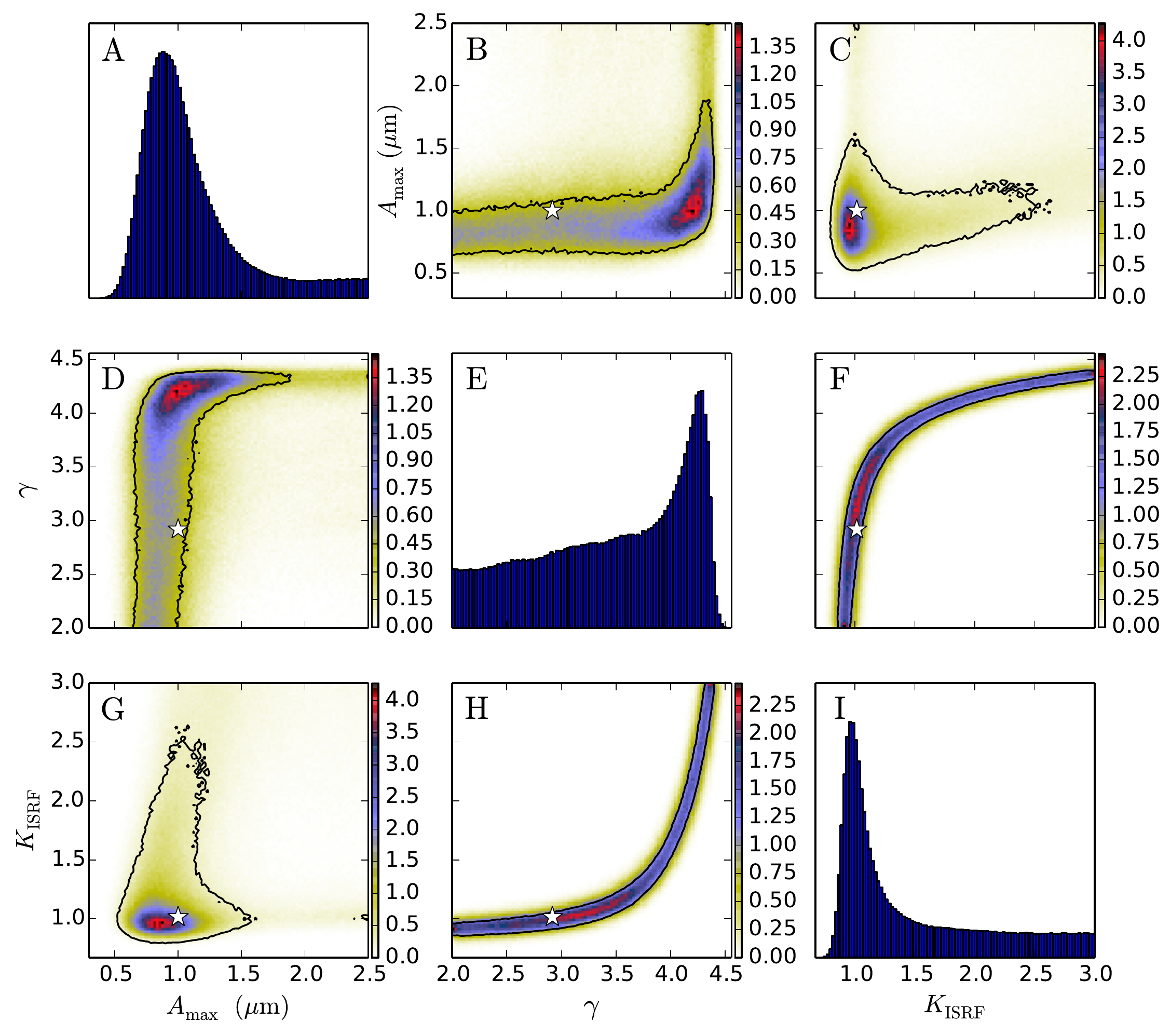} 
\caption{Marginalised probability distributions of dust parameters for the TMC-1N $\tau_J =  6$ position. For the computations, only the J, H, and K bands were used. The white star indicates the projected position of the $\chi^2$ minimum and the black contours show the 1 $\sigma$ of the projection. The colour scale shows the normalised probability.}
\label{fig:MCMC_JHK}
\end{figure*}

When only J, H, and K band observations are used, $A_{\rm max} = 0.8$ $\mu$m and $\gamma=3.2$ (Fig. \ref{fig:MCMC_JHK}). The probability of $A_{\rm max} > 2.0$ $\mu$m has also decreased considerably.

\begin{figure*}
\sidecaption
\includegraphics[width=12cm]{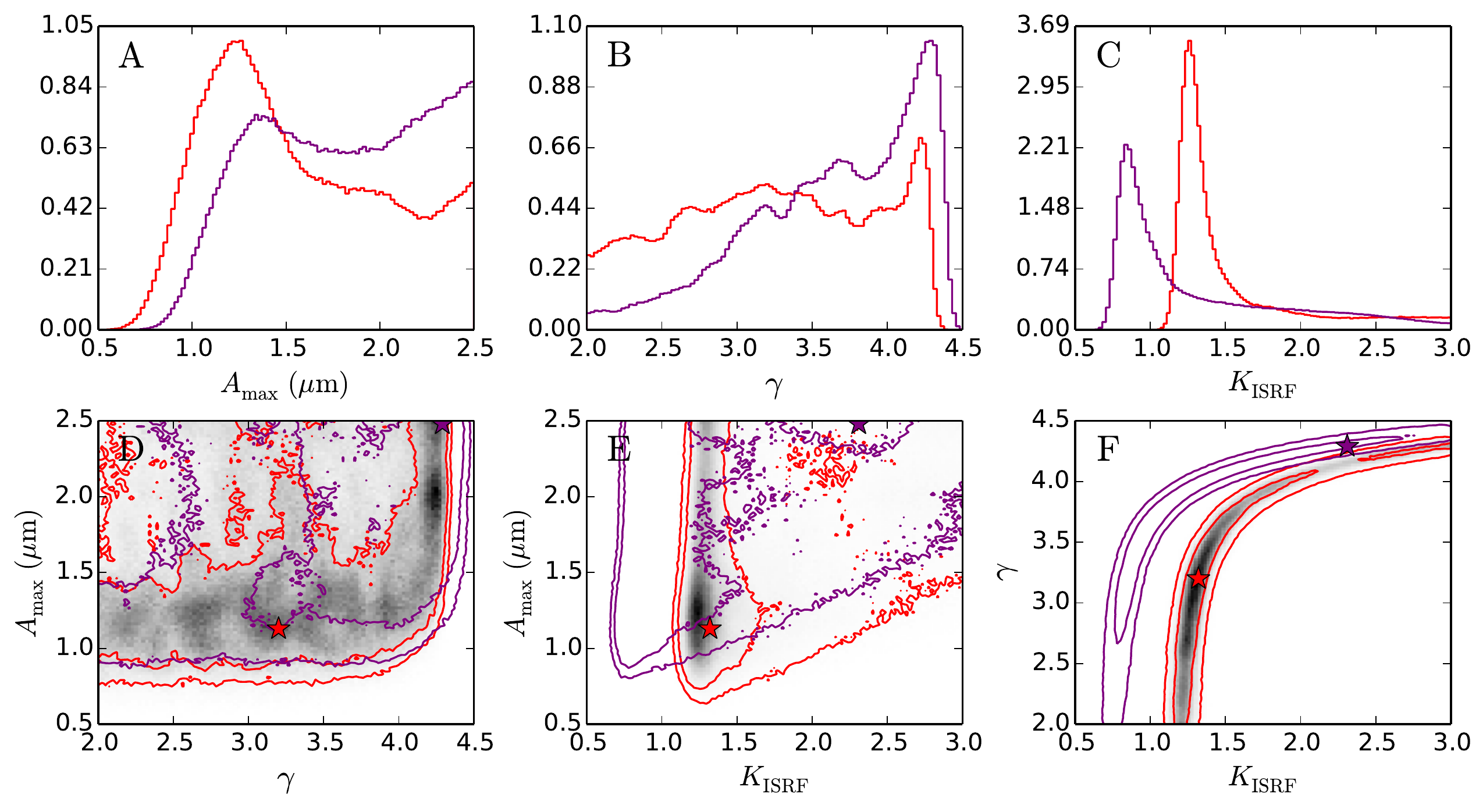} 
\caption{As Fig. \ref{fig:centre_control_10_20}, but the grain albedo in the radiative transfer calculations is changed by -10\% (red lines) or +10\% (purple lines), compared to Fig. \ref{fig:centre_control_10_20}. The stars indicate the projected positions of the $\chi^2$ minima. The grey-scale map corresponds to the red lines and the contours show the 1 and 2 $\sigma$.}
\label{fig:centre_albedo_m_p_10}
\end{figure*}

According to Fig. \ref{fig:centre_albedo_m_p_10}, decreasing the albedo by $10\%$ produces a sharper distribution around $A_{\rm max}$ $\sim 1$ $\mu$m, and increases the probability of small grain sizes, $A_{\rm max} < 1.3$ $\mu$m and $\gamma < 3.6$. A 10\% higher albedo will decrease the estimated ISRF strength by $\sim 20 \%$. Furthermore, a higher albedo will also increase the distance between $\chi^2$ minimum and the peak of the marginalised probability. A larger $\sigma_{\rm NIR}$ will result in broader and flatter parameter distributions (Fig. \ref{fig:centre_albedo_m_p_20}).

Varying the asymmetry parameter $g$ by $\pm$ 10\% produces similar results. Although, the projected position of the $\chi^2$ minimum is shifted relative to the maxima of the marginalised probability distributions (\ref{fig:centre_control_10_20}). With larger $\sigma_{\rm NIR}$, the probability distributions are again wider and flatter and their maxima match better with the $\chi^2$ minima.

\subsubsection{MCMC results, $\tau_J = 2$ position}\label{Sect:3.5}

\begin{figure*}
\sidecaption
\includegraphics[width=12cm]{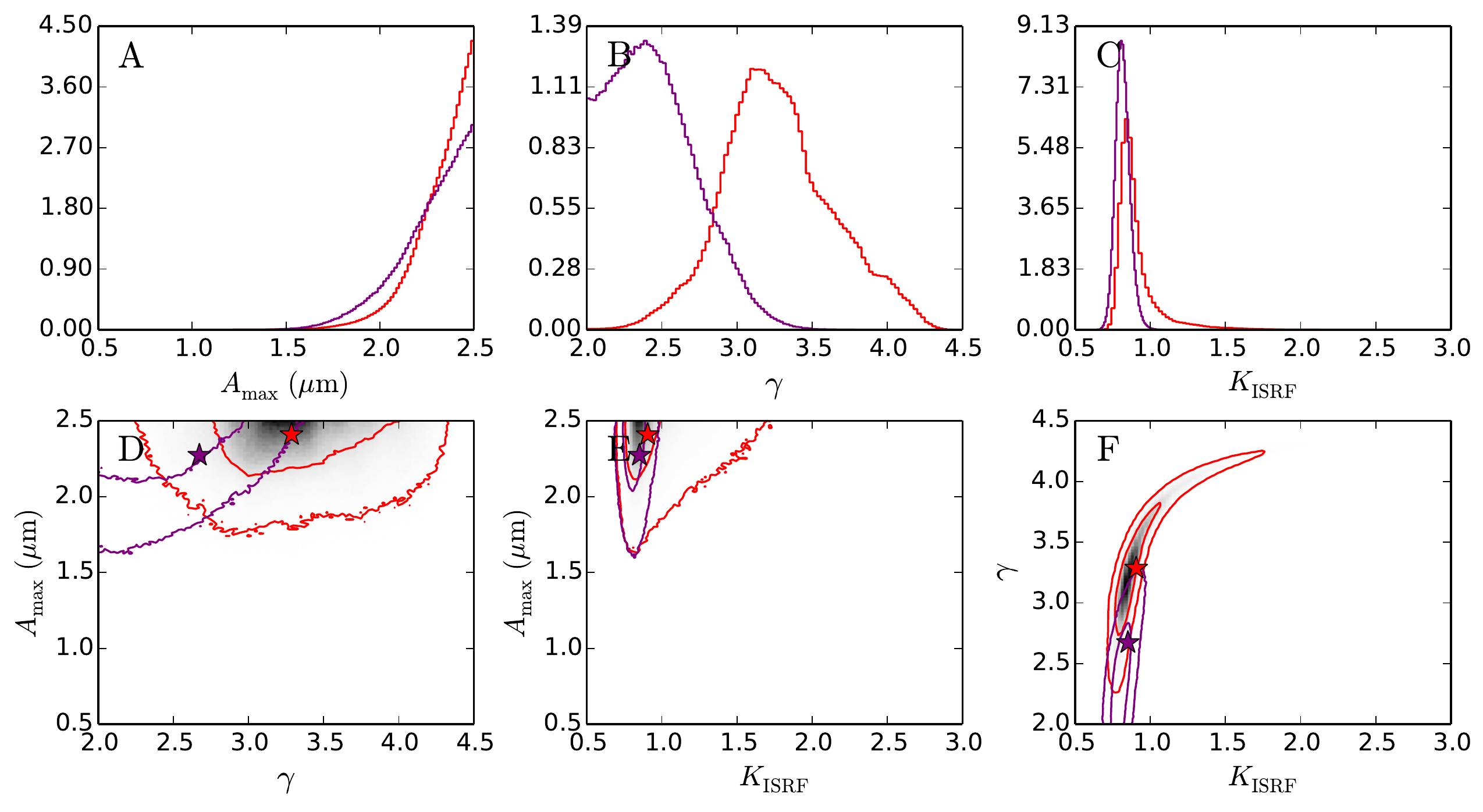} 
\caption{Marginalised probability distributions of dust parameters for the TMC-1N $\tau_J = 2$ position assuming $\sigma_{\rm NIR} = 10\%$, red lines, and $\sigma_{\rm NIR} = 20\%$, purple lines. The stars indicate the projected positions of the $\chi^2$ minima. The grey-scale map corresponds to the red lines and the contours show the 1 and 2 $\sigma$.}
\label{fig:near_control_10_20}
\end{figure*}

Shown in Fig. \ref{fig:near_control_10_20} are the results of the MCMC computations of the TMC-1N $\tau_J = 2$ position. With lower optical depth the $A_{\rm max}$ distribution is limited from above only by the imposed upper limit $A_{\rm max} = 2.5$. Probably because of the same reason $\gamma$ values are all concentrated around 3.5. Compared to TMC-1N $\tau_J =  6$ position, the ISRF estimate is lower with $K_{\rm ISRF} \sim 1.0$.

Assuming larger uncertainties for the surface brightness data has a strong effect (Fig. \ref{fig:near_control_10_20} B) shifting the distribution from $\gamma \sim 3.0$ to $\gamma \sim 2.4$, and decreasing $K_{\rm ISRF}$ from $\sim 1.0$ to 0.8. This suggests some tension between the observations and the model, since for a more diffuse line-of-sight the dust grains should be less evolved with smaller $A_{\rm max}$ and higher $\gamma$. In contrast, there is almost no effect on the $A_{\rm max}$, with the exception of slight broadening of the distribution. The lower optical depth has led to smaller offsets between $\chi^2$ minima and the maxima of projected probability.

The computations carried out by varying the albedo of the dust grains, Fig. \ref{fig:near_albedo_g_10}, produce similar effects as in Fig. \ref{fig:centre_albedo_m_p_10}. Decreasing the albedo increases the required radiation field strength and increases the probability of large grains. Changing the asymmetry parameter does not effect the derived parameter distributions.

\subsubsection{Grain composition}\label{Sect:3.6}

In the previous sections, we have used models that have equal J band optical depth for silicate and carbon grains. In order to study how the chemical composition of the grains affects the scattering properties we examine two alternative dust models with $r_{\rm Si}$=0.7 and $r_{\rm Si}$=0.3 ratios.

\begin{figure*}
\sidecaption
\includegraphics[width=12cm]{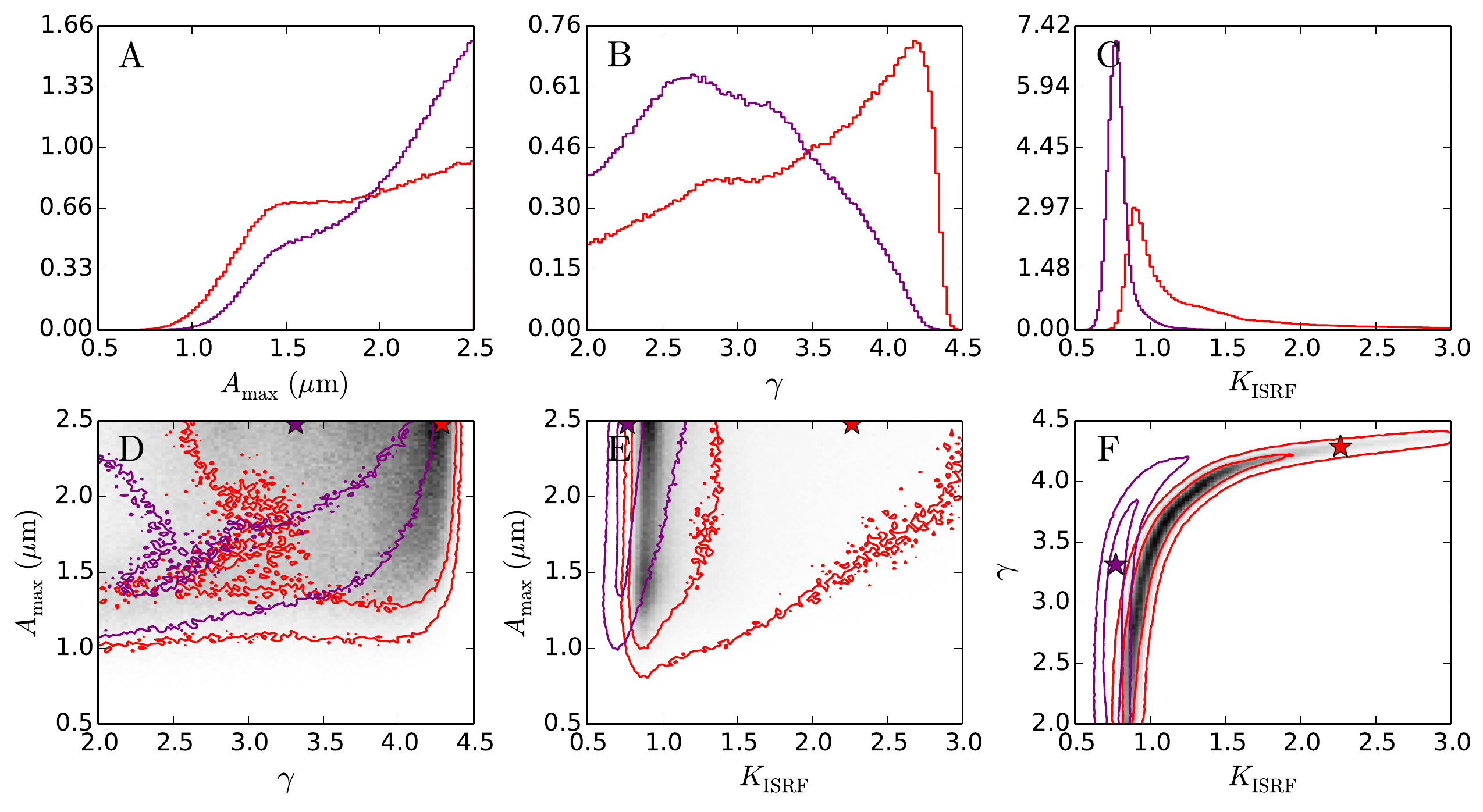} 
\caption{Marginalised probability distributions of dust parameters for the TMC-1N $\tau_J =  6$ position with a $r_{\rm Si}$=0.3, red lines, and $r_{\rm Si}$=0.7, purple lines, respectively. For both computations $\sigma_{\rm NIR} = 10\%$ is assumed. The stars indicate the projected positions of the $\chi^2$ minima. The grey-scale map corresponds to the red lines and the contours show the 1 and 2 $\sigma$.}
\label{fig:carbon_silicate_10}
\end{figure*} 

The model with $r_{\rm Si}$=0.7 prefers a lower value of $\gamma$ than the model with $r_{\rm Si}$=0.3 (Fig. \ref{fig:carbon_silicate_10}). The shapes of the two distributions could explain the broad $\gamma$ distribution seen in the model computed with synthetic observations (Fig. \ref{fig:center_simu_10}). With $r_{\rm Si}$=0.7 the $\gamma$ distribution is skewed towards $\gamma < 3.5$, but with $r_{\rm Si}$=0.3 the distribution is skewed towards $\gamma > 3.5$, thus, a model with $r_{\rm Si}$=0.5 will have a broader $\gamma$ distribution.

The $A_{\rm max}$ distributions of both models have rather similar shapes, with a plateau at $\sim1.5\mu$m, although the distribution is broader for the $r_{\rm Si}$=0.3 case. The model with $r_{\rm Si}$=0.7 produces a sharp and narrow distribution around $K_{\rm ISRF}$ $\sim 0.7$, whereas the computation with $r_{\rm Si}$=0.3 is centred around $K_{\rm ISRF}$ $\sim 1.0$, with a tail towards higher values. The differences can be explained by carbon having a higher overall absorption efficiency, whereas silicates have higher scattering efficiency. With $r_{\rm Si}$=0.3, the $\chi^2$ minimum does not match peak of the marginalised probability.

\section{Discussion}\label{Sect:4}

We have used radiative transfer computations to study how changes in the properties of interstellar dust grains, maximum grain size and size distribution, affect the observed scattered light at NIR and MIR wavelengths. We have shown that the scattered light can be used to constrain dust parameters. In our radiative computations we have assumed that the observed surface brightness consists of scattered radiation and of background radiation seen through the cloud. In particular, we assume that the thermal emission is negligible. We use a spherical one-dimensional cloud model and a three-dimensional ellipsoid model which are illuminated by an isotropic interstellar radiation field, or by an anisotropic radiation field. The properties of the dust grains are based on the model by \citep{WD01}.

In our simulations we have only used one of the Spitzer bands, but, as discussed by \citet{Lefevre2016}, light scattering even at 8 $\mu$m may not be negligible. Careful modelling of all Spitzer bands, or even longer wavelengths for example using far-infrared emission, can be used to place additional constraints on the dust parameters.

\subsection{Constraints on dust parameters}\label{Sect:4.2}

Based on our simulations, strong correlations between the dust parameters are evident as changes in one parameter affect the other parameters. The $K_{\rm ISRF}$ and $\tau_{\rm J}$ can not be constrained simultaneously, as changes in the optical depth can be compensated by changes in the strength of the radiation field. Similarly, increasing the $K_{\rm ISRF}$ will be shift the $\gamma$ distribution to lower values, as in many of our simulations both parameters tend to have long tails. On the other hand, the colour of the NIR-MIR spectra is affected by changes in the $A_{\rm max}$ and $\gamma$. Furthermore, assumptions of the morphology of the cloud can change the estimated dust parameters significantly.

To categorise the parameter distributions, we use relative uncertainty

\begin{equation}
r (x) = \frac{Q_3 - Q_1}{1.349 \times Q_2},
\end{equation}
where $x$ is the dust parameter, $Q_1$ and $Q_3$ are values of the first and third quartile, $Q_2$ is the median, and the factor 1.349 is the scaling between the interquartile range $Q_3-Q_1$ and the standard deviation in case of a normal distribution.

For the probability distributions that are derived using the synthetic observations, we define the difference between the median of the distribution and the correct value of the dust parameter

\begin{equation}
\Delta (x) = \frac{Q_2 - x}{x}.
\end{equation}

In the computations with the synthetic observations, the probability distributions of $A_{\rm max}$ and $K_{\rm ISRF}$ have a clear peak near the correct values used to derive the observations. For the model shown in red in Fig. \ref{fig:center_simu_10}, the $\Delta (A_{\rm max}) = 13\%$ and $r(A_{\rm max}) = 25 \%$ and the $\Delta (K_{\rm ISRF}) = -1.7\%$ and $r (K_{\rm ISRF}) = 61 \%$. However, the $\gamma$ distribution of the model does not have a clear peak around the true value, however, the resulting $\Delta (\gamma) = -0.5 \%$ and $r (\gamma) = 24 \%$. The difference between assuming a major axis or minor axis radiative transfer model can change the estimated amount of large grains, by changing the $A_{\rm max}$ by up to $\sim 30\%$ and $\gamma$ by up to $\sim 10\%$ and affect the estimated $K_{\rm ISRF}$ by up to $\sim 30\%$.

Small variations in the optical properties of the dust grains are reflected to the derived dust parameters. If the albedo of the grains is increased by $10 \%$ in the scattering simulations the probability of larger grain sizes increases as seen in Fig. \ref{fig:simu_albedo_mp}. The peak of the $A_{\rm max}$ distribution shifts from $A_{\rm max} \sim 1$ $\mu$m up to $A_{\rm max} \sim 1.5$ $\mu$m, and the probability of grain sizes in excess of 2 $\mu$m increases. However, the increase in albedo in the simulations also increases the probability of $\gamma > 3.5$ and decreases the strength of the required radiation field. Thus, although the maximum grain size increases, the relative amount of large grains is decreased. Similarly the $\Delta (A_{\rm max}) = 27 \%$ and $r (A_{\rm max}) = 26 \%$, whereas for the $\gamma$ distribution the $\Delta (\gamma) = 5.2 \%$ and $r (\gamma) = 18 \%$ and for the $K_{\rm ISRF}$ distribution the $\Delta (K_{\rm ISRF}) = 5.8 \%$ and $r (K_{\rm ISRF}) = 71 \%$. Decreasing the albedo by $10 \%$ results in deviations between the correct values and the median values of $\Delta (A_{\rm max}) = 1.4\%$, $\Delta (\gamma) = -0.8\%$, and $\Delta (K_{\rm ISRF}) = 14 \%$ and the relative uncertainties are $r (A_{\rm max}) = 23 \%$, $r (\gamma) = 28 \%$, and $ r (K_{\rm ISRF}) = 38 \%$.

\begin{figure}
\resizebox{\hsize}{!}{\includegraphics[width=17cm]{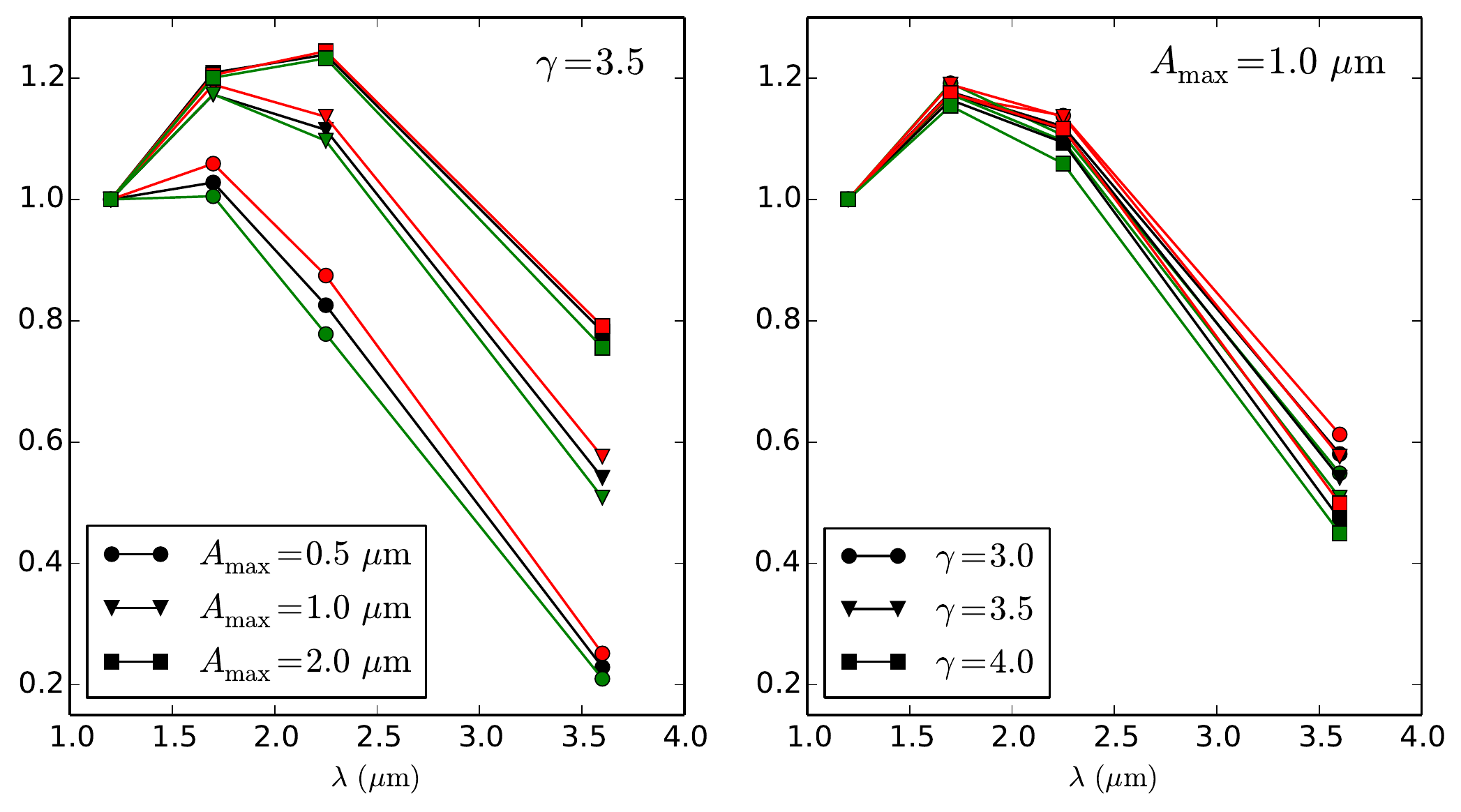}} 
\caption{Near-Infrared spectra derived from different dust models, for the left panel a constant $\gamma =3.5$ is used and for the right panel $A_{\rm max} = 1.0 \mu$m. The red and green lines are from models where the albedo of the dust grains has been changed during the scattering computations by $-10 \%$ (red) or by $+10 \%$ (green) and the black line is from a model where only the $A_{\rm max}$ and $\gamma$ were varied during the scattering computations. The symbols correspond to different $A_{\rm max}$ and $\gamma$ values in the left and right panels, respectively.}
\label{fig:albedo_spec}
\end{figure}

The changes in the shape of the $A_{\rm max}$ and $\gamma$ distributions are related to the way radiation is scattered by the grains. A higher albedo decreases the absorption efficiency of the dust grains resulting in a spectra that is bluer compared to the reference spectra, thus, larger grain sizes are needed. However, because of the increased scattering efficiency, the amount of large grains is smaller. Shown in Fig. \ref{fig:albedo_spec} are the spectra computed from three different dust models by varying either the $A_{\rm max}$ or $\gamma$ parameter. Changes in the $A_{\rm max}$ have stronger effect on the resulting intensities compared to changes in $\gamma$.

An error of 10\% in the intensity of a single band (see Fig. \ref{fig:center_simu_10}) affects the derived value of $A_{\rm max}$ by $< 10 \%$, but, the derived value of $K_{\rm ISRF}$ can change by up to 60\%. Uncertainty in the observed H or K band intensity affects the derived parameter values more than changes in J or 3.6 $\mu$m bands.

Many of our MCMC computations show a high probability for grain sizes in clear excess of 2 $\mu$m, but the estimated size distribution can also be changed by using different type of grains. We have used a simple model with bare silicate and carbon grains, thus, in order to increase the surface brightness of scattered light at longer wavelengths the maximum grain size has to be increased. However, the scattering properties of the dust grains can be changed by including ice mantles, coagulation, or by changing the surface composition (e.g. hydrogen rich carbon mantles) of the grains, thus increasing the scattering efficiency without increasing the grain sizes \citep{Ysard2016,Jones2016}.

\begin{figure*}
\sidecaption
\includegraphics[width=12cm]{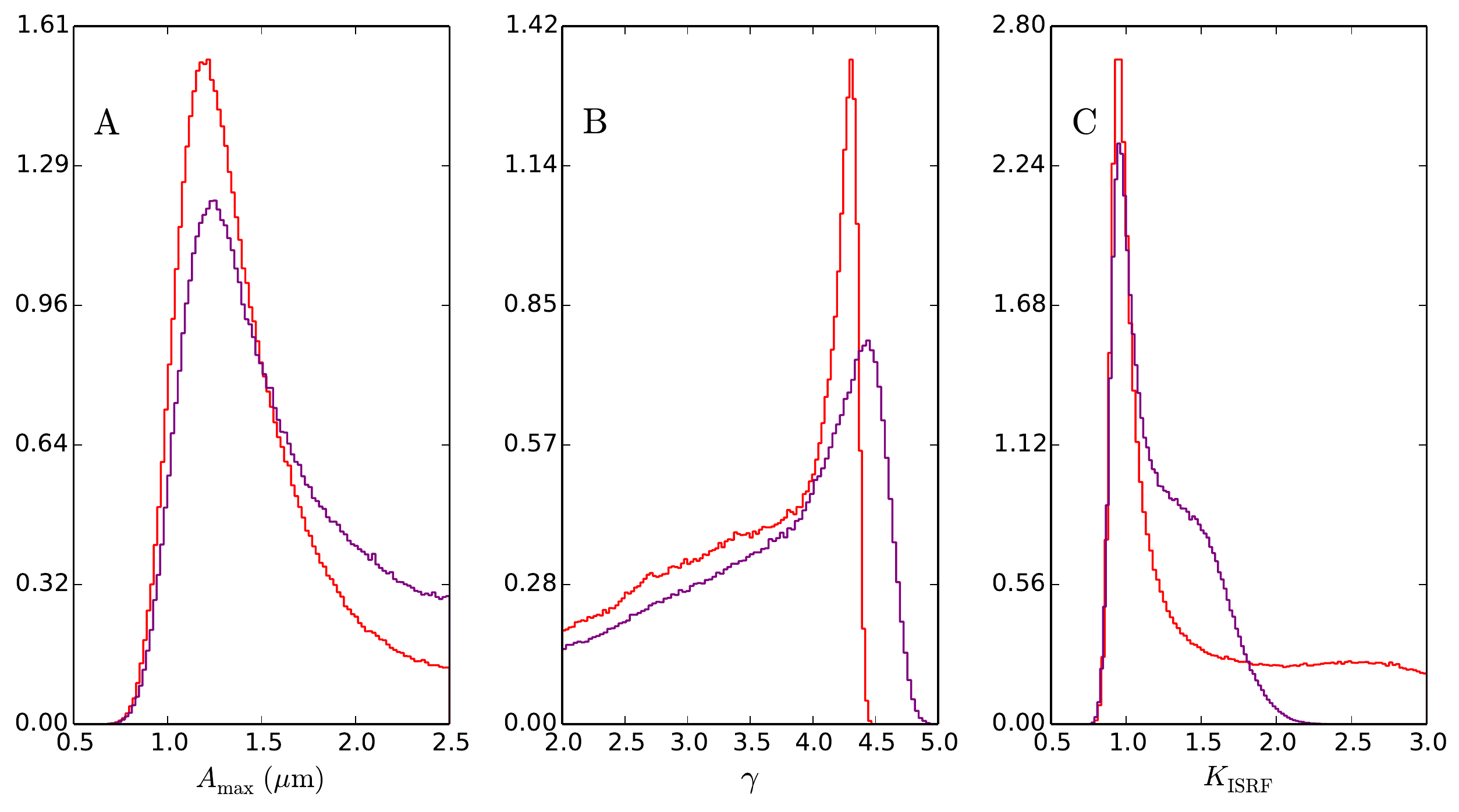}
\caption{Marginalised probability distributions of the dust parameters for the minor axis case, red lines (see Fig. \ref{fig:simu_control}), and the case with grains smaller than 20 nm removed from the dust model used to fit the synthetic observations, purple lines.} 
\label{fig:nosmall}
\end{figure*}

Coagulation and formation of ice mantles will change the size distribution, decreasing the amount of small grains. As seen in \ref{fig:gra_sil}, the small grains can cause significant amount of absorption, and thus, reduce the amount of photons available for scattering. We have run separate computations, where we have removed all grains smaller than 20 nm from the dust model. The results of the simulations, using the synthetic observations of the minor axis case (see Fig. \ref{fig:simu_control}) and assuming $\tau_{\rm J} = 6$, are shown in Fig. \ref{fig:nosmall}. Compared to the minor axis case (red lines), removing small grains from the model that is used to fit the synthetic observations (purple lines) does have an effect on the resulting marginalised probability distributions but it is not large. The peak of the $\gamma$ distribution is shifted to a higher value of $\gamma = 4.5$ and the $K_{\rm ISRF}$ distribution has a upper limit of $K_{\rm ISRF} = 2$.

\begin{figure*}
\sidecaption
\includegraphics[width=12cm]{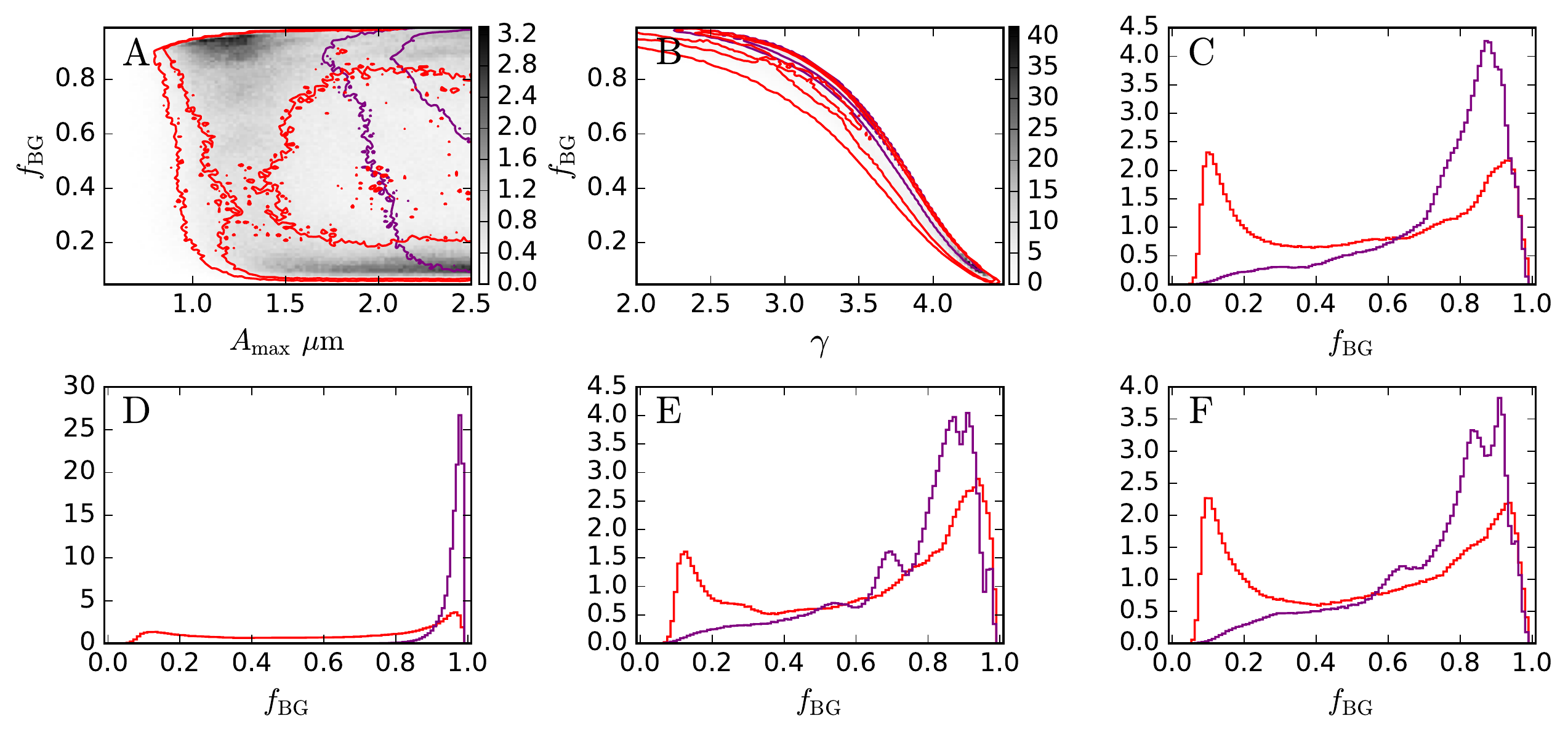} 
\caption{The mass fraction of grains larger than 0.25 $\mu$m for the cases $\tau_J =  6$ (red lines) and $\tau_J =  2$ (purple lines). For panels A-C, the $A_{\rm max}$ and $\gamma$ values of the model with $r_{\rm Si}$=0.5 and $\sigma_{\rm NIR} = 10\%$ were used. For panel D, we use the previous model, but assume $\sigma_{\rm NIR} = 20\%$. For panels E and F, we use the models with the albedo of the grains decreased by $10 \%$ and the asymmetry parameter decreased by $10 \%$, respectively. The gray scale colour maps correspond to the red lines and the contours show the 1 and 2 $\sigma$.}
\label{fig:dustmass}
\end{figure*}

The changes in $A_{\rm max}$ and $\gamma$ affect the fraction of mass in large grains (Fig. \ref{fig:dustmass}) 

\begin{equation}
f_{\rm BG} = \frac{m(a > 0.25 \mu \rm m)}{m},
\end{equation}
where $m(a > 0.25 \mu \rm m)$ is the mass of grains larger than 0.25 $\mu$m and $m$ is the total mass. In all cases, the mass fraction in large grains increases sharply when the value of $\gamma$ decreases. In the case of lower optical depth, the $f_{\rm BG}$ is higher compared to the $\tau_{\rm J} = 6$ case. However, based on the MCMC results, the probability that a diffuse line-of-sight would have more large grains than a optically thicker line-of-sight is $p(f_{\rm BG}(\tau_{\rm J} = 6) > f_{\rm BG}(\tau_{\rm J}= 2)) = 25\%$. The result is not statistically meaningful, thus, the evidence for a high amount of large grains towards the $\tau_{\rm J}=2$ line-of-sight is not very strong.

In many of our computations, the estimated $K_{ISRF}$ has a clear peak around $\sim 1.1$, but $K_{\rm ISRF}$ could also be significantly higher. However, if the radiation field and optical depth can be constrained, for example setting a constant value for $K_{\rm ISRF}$ and $\tau_{\rm J}$ (Fig. \ref{fig:no_isrf}) results in well constrained parameter distributions for both $A_{\rm max}$ and $\gamma$ with $\Delta (A_{\rm max}) = 3.4 \%$, $r (A_{\rm max}) = 20 \% $ and $\Delta (\gamma) = 0.2 \%$, $r (\gamma) = 3.2 \%$. Assuming a prior 20\% uncertainty in $K_{\rm ISRF}$ and $\tau_{\rm J}$, results in $A_{\rm max}$ and $\gamma$ distributions that are broader ($\Delta (A_{\rm max}) = 4.2 \%$, $r (A_{\rm max}) = 18 \%$ and $\Delta (\gamma) = 7.9 \%$, $r (\gamma) = 17 \%$, respectively).

The analysis of the computations where the synthetic observations are used shows that on average the $r (A_{\rm max}, \gamma) \sim 25\%$, whereas the $r (K_{\rm ISRF})$ is $\sim 45 \%$ on average. However, the $r (K_{\rm ISRF})$ can reach values as high as $\sim 70\%$.

\begin{figure*}
\sidecaption
\includegraphics[width=12cm]{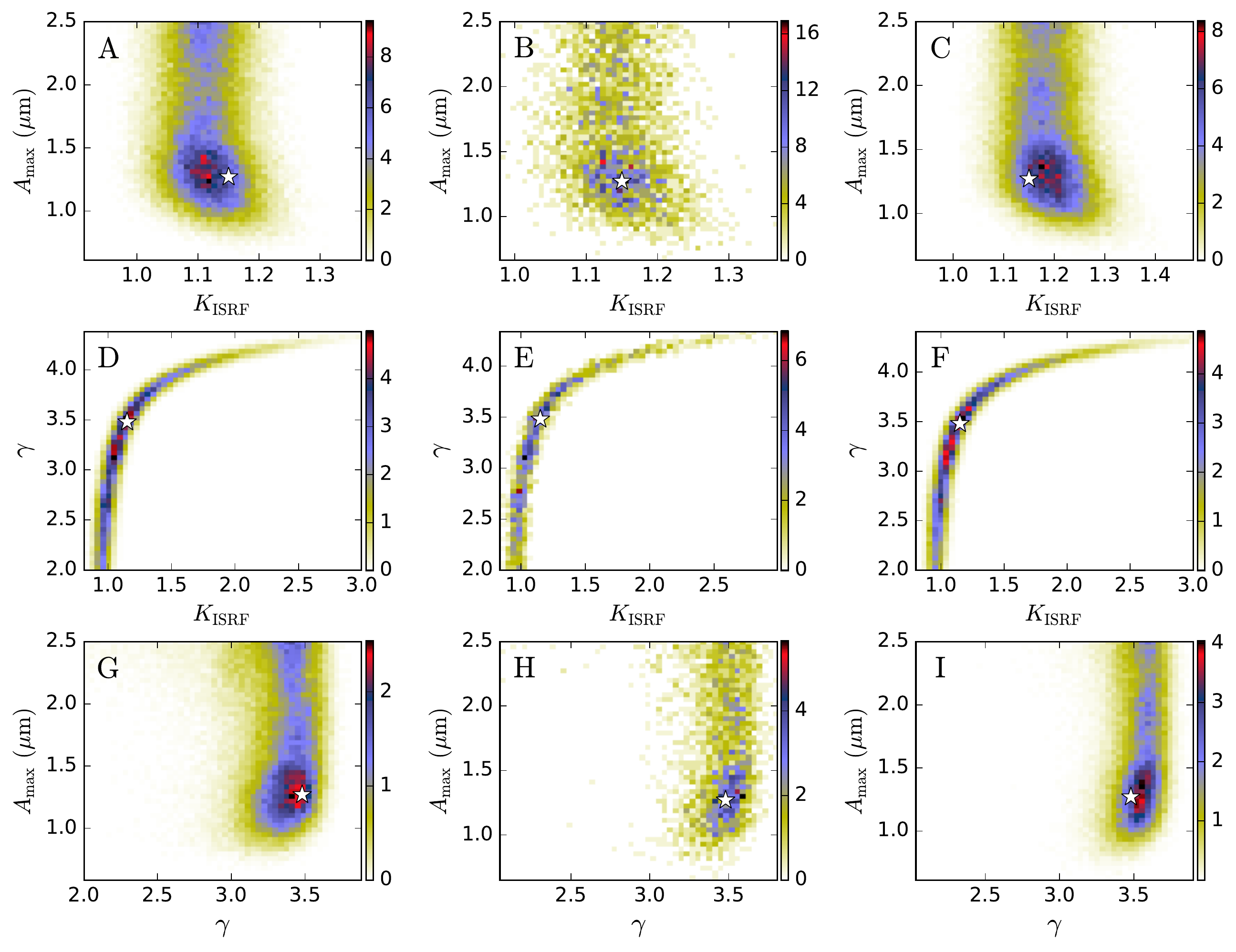} 
\caption{Marginalised probability distributions of dust parameters for the TMC-1N $\tau_J =  6$ position. The white star indicates the projected position of the $\chi^2$ minimum.  The panels A, D, G and C, F, I show the marginalised probability integrated along the third parameter axis within $\pm 5 \%$ of the $\chi^2$ minimum, respectively. Panels B, E, and H show the marginalised probability on the $\chi^2$ minimum.}
\label{fig:xsection}
\end{figure*}

As seen in the synthetic observations, the optical depth of the cloud is not well constrained by studying only the scattered light. This is related to the saturation of the surface brightness, and the degeneracy between radiation field intensity and dust opacity as changes in opacity can be negated with changes in the strength of the radiation field. When possible, analysis should include external constraints on the opacity or the radiation field that can be obtained from dust emission or directly from extinction measurements.

If the radiation field strength and the optical depth of the cloud can be constrained, the surface brightness ratio between J and K bands can be used to place useful constraints on the dust properties. However, for diffuse regions, the location of the cloud with respect to the source of illumination should be taken into account since differences of the asymmetry parameter can change the relative brightness between the J and K bands, however, the effect of the asymmetry parameter also depends on the optical depth of the cloud.

\subsection{Limitations of the models}

We have assumed that the background in Eq. \ref{eq:sironta_malli} is truly a background and there is no emission from between the cloud and the observer. Emission originating between the cloud and the observer causes error in the assumed transmitted radiation $I_{\rm bg} e^{-\tau}$ and thus also in the deduced colour of the scattered component. A more detailed radiative transfer study is needed to explore the effects arising from the surrounding medium. In our simulations we have used smooth spherical and ellipsoidal cloud models but have not considered effects of small-scale density inhomogeneities that could affect the way radiation penetrates the cloud. Furthermore, we have used a constant scaling for the strength of the radiation field $K_{\rm ISRF}$, thus we have not considered possible changes in the shape of the background spectra.

The $K_{\rm ISRF}$ distribution tends to have a long tail towards higher values, while the powerlaw exponent of the grain size distribution $\gamma$ has a tail towards lower values. The tails are probably caused by the chosen prior distributions of the parameters. Allowing the $K_{\rm ISRF}$ to have values larger than 3.0, would likely allow the $\gamma$ distribution to reach even smaller values.

The position of the $\chi^2$ minimum does not usually coincide exactly the maximum of the marginalised probability. The three dimensional function $\chi^2(A_{\rm max}, \gamma ,K_{\rm ISRF})$ is complex, and the minimum is shallow. Furthermore, the values of the $\chi^2$ function are computed from simulations that have noise, thus the actual location of the computed $\chi^2$ minimum can be affected by noise.

The marginalised probability distributions are projections of a three dimensional distribution. Thus it is possible that the $\chi^2$ minimum of the three dimensional distribution does not match with the maximum of the projected two dimensional probability. Furthermore, using higher error estimates for the surface brightness data can have a significant effect on the location of the $\chi^2$ minima. The instability of the $\chi^2$ minima, seems analogous with the color temperature and spectral index, $T_{\rm C}$-$\beta_{\rm spec}$, plane. As discussed by \citet{juvela2012}, even a slight change in the weighting of the frequency points or in the noise can significantly shift the location of the global minimum. High uncertainties make the $\chi^2$ minima shallow and, together with noise, contribute to ambiguity in the actual location of the $\chi^2$ minima.

\subsection{Taurus observations}

As an example we have analysed observations of a filament in the Taurus molecular cloud, TMC-1N. Our cloud model is an ellipsoid with a $3:1$ ratio between the major and minor axes, illuminated by an anisotropic radiation field derived from the DIRBE observations and viewed along one of the minor axes. The optical depth of the model cloud along the line-of-sight is set to $\tau_{\rm J} = 6$ or $\tau_{\rm J} = 2$. Based on our analysis for the TMC-1N $\tau_J =  6$ position, a maximum grain size of $A_{\rm max} \sim 1.2$ $\mu$m and $\gamma > 4$ are most probable, with an average relative uncertainty of $r (A_{\rm max}) = 36\%$, $r (\gamma) = 22\%$, and $r (K_{\rm ISRF}) = 38\%$. However, there is a considerable possibility of grains larger than $A_{\rm max} > 2.0$ $\mu$m. The mass weighted average grain size is $\langle a_{\rm m} \rangle = 0.113$, which is higher than $\langle a_{\rm m} \rangle = 0.089$ of the MRN distribution. The anisotropic radiation field derived from the DIRBE observations is sufficient to reproduce the observed surface brightness values.

Shown in Fig. \ref{fig:xsection}, is a cut of the marginalised probability distribution around the $\chi^2$ minimum for the TMC-1N $\tau_J =  6$ position with $r_{\rm Si}$=0.5 and assuming $\sigma_{\rm NIR} = 10\%$, see Fig. \ref{fig:centre_control_10_20}. In all of the panels, the $\chi^2$ minimum matches with the maximum of the marginalised probability.

If one assumes larger observational uncertainties $\sigma_{\rm NIR} = 20 \%$, the resulting parameter distributions are smoother and broader, with an average relative uncertainty of $r (A_{\rm max}) = 26\%$, $r (\gamma) = 24\%$, and $r (K_{\rm ISRF}) = 30\%$. However, the increased $\sigma_{\rm NIR}$, of both TMC-1N $\tau_J =  6$ and $\tau_J = 2$ positions, can also change the interpretation of the dust properties. For example, with a lower $\sigma_{\rm NIR}$, the $A_{\rm max}$ distributions of Fig. \ref{fig:centre_control_10_20} shows a clear peak around $\sim 1$ $\mu$m, and a tail towards higher parameter values and a $r = 37 \%$. Adopting a higher $\sigma_{\rm NIR}$ shifts the peak of $A_{\rm max}$ distribution to higher value, $\sim 1.5$ $\mu$m, causes a significant increase for $A_{\rm max} > 1.5$ $\mu$m, and decreases $r (A_{\rm max})$ to $25\%$. With lower uncertainties the $\gamma$ distribution has a plateau or a bump around 3.5 and the $r (\gamma) = 22\%$, whereas with higher $\sigma_{\rm NIR}$ the distribution is smoother and the $r (\gamma) = 24 \%$.

The 3.6 $\mu$m band will place more constraints on the probability distributions, significantly increasing the probability of $A_{\rm max} > 1.5 \mu$m with $r (A_{\rm max}) = 30\%$, $r (\gamma) = 25\%$, and $r (K_{\rm ISRF}) = 56\%$ However, caution should be taken with the 3.6 $\mu$m band. To derive reliable estimates for the dust parameters the fraction of emission in the observed surface brightness should be know to a reasonable accuracy.

With lower optical depth, case $\tau_{\rm J} = 2.0$, the parameter distributions are significantly narrower with average $r (A_{\rm max}, \gamma, K_{\rm ISRF}) < 10 \%$. The $\gamma$ values are smaller than $\sim 3.8$ and the values of $A_{\rm max}$ are concentrated to values larger than 2 $\mu$m. However, the opposite would be expected, since for a more diffuse line-of-sight, the dust grains are thought to be less evolved, and thus, smaller values of $A_{\rm max}$ and larger values of $\gamma$ would be expected. Increasing the $\sigma_{\rm NIR}$ pushes the $\gamma$ distribution to lower value $\gamma \sim 2.4$. The change in the $\gamma$ distribution is likely related to the shape of the $\chi^2$ space and changes in the assumed error estimates can also strongly affect the way the priori affect the posterior probability distributions.

The $K_{\rm ISRF}$ distribution is slightly lower in the case $\tau_{\rm J} = 2.0$, although the radiation field there should be less attenuated. A region with lower optical depth could be used to constrain the strength of the radiation field. For case $\tau_{\rm J} = 2.0$, the intensity of the 3.6 $\mu$m band is low, which introduces additional uncertainty to the derived dust parameters.

\section{conclusions}\label{Sect:5}

\renewcommand{\labelitemi}{$-$}

We have used radiative transfer computations of near-infrared scattered light with Markov chain Monte Carlo method to study the optical properties and size distributions of interstellar dust. We have examined how NIR (MIR) observations of scattered light can be used to constrain properties of the dust grains and the radiation field illuminating dense clouds.

The main results of our study are
\begin{itemize}

\item Our analysis of the synthetic observations indicates that for surface brightness measurements with $\sim 10\%$ errors, the relative uncertainty of the main dust parameters $A_{\rm max}$ and $\gamma$ are $\sim 25\%$. The relative uncertainty of $K_{\rm ISRF}$ is higher $\sim 45\%$. The difference between the median of the posterior probability distribution and the correct parameter value is typically $< 15\%$.

\item Tests with ellipsoidal models with aspect rations $1:3$ showed that the assumed morphology of the cloud can change the estimated $K_{\rm ISRF}$ and $A_{\rm max}$ by up to $\sim 30\%$ and the estimated value of $\gamma$ by $\sim 10\%$.

\item An error of 10\% in the surface brightness of one channel does not significantly affect the derived value of $A_{\rm max}$. However, the uncertainty affects the derived values of $\gamma$ and $K_{\rm ISRF}$ by up to 30\% and 60\%, respectively, depending on which of the wavelength bands has the uncertainty. Uncertainty in the H or K band has stronger effect on the derived dust parameters than J or 3.6 $\mu$m band, as they affect the shape of the spectra more.

\item The strength of the radiation field $K_{\rm ISRF}$ and the optical depth $\tau_{\rm J}$ can not be well constrained simultaneously. However, other independent methods can be used to derive constraints for $\tau_{\rm J}$, for example by extinction studies. If the $K_{\rm ISRF}$ and $\tau_{\rm J}$ are known a priori to an accuracy of $\sim 20 \%$, the uncertainty of $A_{\rm max}$ and $\gamma$ is decreased to $\sim 18\%$. 

\item The prior used in our MCMC computations cause long tails in the $\gamma$ and $K_{\rm ISRF}$ distributions.

\item In our tests, when synthetic observations were analysed with radiative transfer models with $10\%$ higher albedo the $A_{\rm max}$ and $\gamma$ distributions shift to higher values, $A_{\rm max} > 1.3$ $\mu$m and $\gamma > 3.5$. However, the required radiation field strength is decreased to $K_{\rm ISRF} \sim 0.8$.

\item The $\chi^2$ minima and the maximum of marginalised probability do not correspond exactly to the used parameter values in most of our computations. The three dimensional function $\chi^2(A_{\rm max}, \gamma ,K_{\rm ISRF})$ is complex, and the minimum of the the $\chi^2$ distribution can be very shallow.

\item Considering the TMC-1N observations, for a line of sight with $\tau_{\rm J} = 6$, a maximum grain size $A_{\rm max} > 1.5$ $\mu$m and a size distribution with $\gamma > 4.0$ have high probability, with an average relative uncertainty of $\sim 25 \%$. However, there seems to be a strong peak around $A_{\rm max} = 1$ $\mu$m. The background radiation field derived from the DIRBE observations is sufficient to reproduce the observed surface brightness with $K_{\rm ISRF} \sim 1$ and the average relative uncertainty is $\sim 30 \%$. The mass weighted average grain size is $\langle a_{\rm m} \rangle = 0.113$, compared to $\langle a_{\rm m} \rangle = 0.089$ of the MRN distribution.

\item We have used a simple dust model by \citet{WD01} that only includes bare silicate and carbon grains. Using a more refined model with additional properties like ice mantles, will change the scattering properties of the grains and can result in smaller grain sizes.

\end{itemize}

\begin{acknowledgements}
MS and MJ acknowledge  the support of the Academy of Finland Grant No. 285769. JMa acknowledges the support of ERC-2015-STG No. 679852 RADFEEDBACK.
\end{acknowledgements}

\bibliographystyle{aa}
\bibliography{Near-infrared_scattering_as_a_dust_diagnostic}

\begin{appendix}

\section{Near infrared observations}

The near-infrared observations in the J, H, and K band, corresponding to 1.25, 1.65, and 2.22 $\mu$m, respectively are from the Wide Field CAMera (WFCAM) instrument of the United Kingdom InfraRed Telescope (UKIRT). The images are centred at RA(J2000) 4h39m36s, DEC(J2000) +26$^{\circ}$39'32'' covering an area of $1^\circ \times 1^\circ$ corresponding to some (2.4 pc)$^2$. The data reduction is described in \citep{Malinen} and we carried out additionally the subtraction of point sources.

The 3.6-$\mu$m and 4.5 $\mu$m band observations are from the Spitzer InfraRed Array Camera (IRAC) \citep{Spitzer}. The images were obtained from the NASA/IPAC Infrared Science Archive (IRSA). The images are centred on the TMC-1N filament in the Taurus molecular cloud complex, the coordinates of the center position are RA(J2000) 4h37m45s, DEC(J2000) +26$^{\circ}$57'24''. The data (observation numbers 11230976 and 11234816) are from the Taurus Spitzer legacy project (PI D. Padgett). We used the level-2 data.

For background subtraction we use images from the DIRBE satellite. The DIRBE images were obtained from the Sky View virtual observatory. For our data analysis, we only needed the first three DIRBE bands which correspond to wavelengths of 1.25, 2.2, and 3.5 $\mu$m, respectively. We used the Zodi-Subtracted-Mission-Average (ZSMA) images, for which the modelled interplanetary dust signal has been subtracted. We have rescaled the DIRBE magnitudes to 2MASS magnitudes; the first two bands were scaled using the filter conversion described by \citet{Levenson} and the third band was scaled using the conversion factors given by \citet{Flagey}.

The DIRBE bands 1, 2, and 3 can be used directly, after the conversion to 2MASS magnitudes, to specify the intensity at the corresponding wavelengths 1.2, 2.2, and 3.5 $\mu$m. The H band intensity was computed by multiplying the average of the J- and K-band intensities with the intensity ratio $I_H / <I_J, I_K>$ derived from the \citet{ISRF} ISRF model.

The background intensities, $I_{\rm BG}$, for J, H, and K bands were adopted from \citet{Malinen}, who used DIRBE and WISE observations to derive the intensities. However, the value of the background intensity of the 3.6 $\mu$m band indicated by Spitzer is higher, 0.1315 MJy/sr, than the value derived by \citet{Malinen}, 0.0765 MJy/sr. Thus, we have used the average of the Spitzer and DIRBE values. The background indicated by Spitzer was obtained by taking an average value over a section in the 3.6 $\mu$m map close to the filament (RA(J2000) 4h41m30s, DEC(J2000) +26$^{\circ}$25'0'' covering an area of $\sim$ $30" \times 30"$) and the value indicated by DIRBE was adopted from \citet{Malinen}. The obtained background intensities are listed in table \ref{tab:observations}. In addition to deriving the background intensities, we have used the DIRBE ZSMA maps to produce an anisotropic radiation field for our radiative transfer computations.

In order to investigate the intensity of the diffuse signal, point sources must be subtracted. We have used both SExtractor \citep{Sextractor} and PSFex \citep{Psfex} to subtract the point sources from the Spitzer and WFCAM images. The subtraction was done in three steps: in the first step, we used the Spitzer and WFCAM images as input for SExtractor. For the initial run, we wanted to detect only the brightest point sources. Thus, we used a high detection threshold corresponding to 15 times the estimated noise in the images. The size of the background mesh was set at 64 pixels and the photometric parameters were set to the correct values for the corresponding Spitzer and WFCAM bands.

With the above parameters, SExtractor produces a catalogue of the brightest point sources and their positions on the image. For the second step, we used the catalogue as input for PSFex, which computes a point-spread function (PSF) and constructs a small model image of the PSF, for each source in the catalogue. To avoid loss of information, PSFex uses a sampling that is different from the original pixels depending on the observations, a rougher sampling for the oversampled and a finer sampling for the undersampled observations. 

In the final step, we used the PSF file and the original images as input of a second SExtractor run. For this run, we used a significantly lower detection threshold, 1.5 times the estimated noise, to detect and later to remove even faint sources. The other parameters discussed above were not changed. The second run produces an image of objects which was subtracted from the original images resulting in an image in which stars appear as smooth holes.

\begin{figure*}
\centering
\includegraphics[width=17cm]{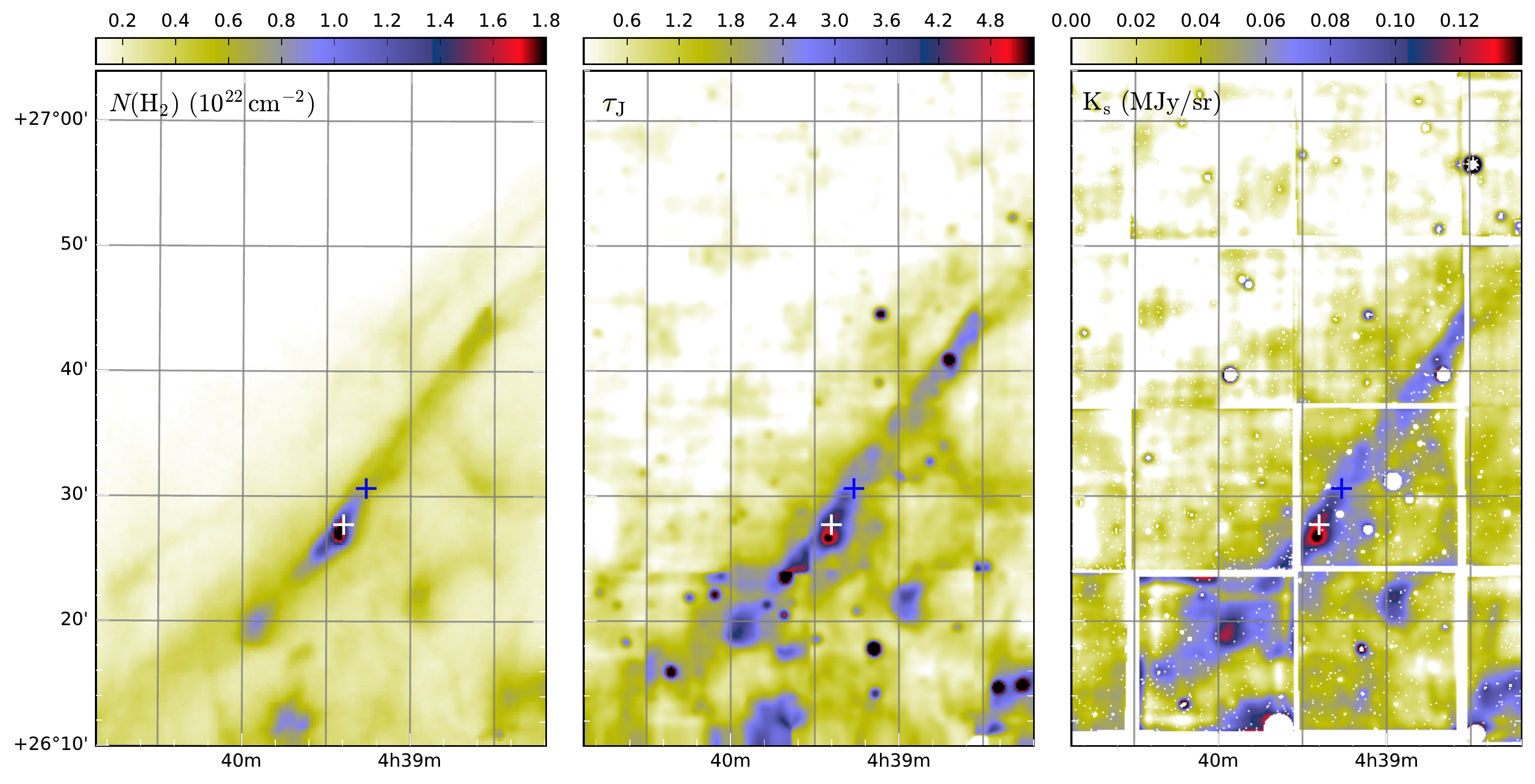} 
\caption{$0.6^{\circ} \times 0.6^{\circ}$ hydrogen column density map derived from the Herschel data, left, and the optical-depth map derived using the surface brightness of the scattered light, center \citep{Malinen}. Shown on the right, a $\rm K_s$ band intensity map (MJy/sr) \citep{Malinen}. The positions used to represent the filament in the CRT and MCMC simulations are shown with a white plus-sign ($\tau_{\rm J} \sim 6$) and with a blue plus-sign ($\tau_{\rm J} \sim 2$).}
\label{fig:vertailu}
\end{figure*}

Map of optical depth at 250 $\mu$m, derived from the Herschel submillimetre observations, and a map of optical depth of the J band, derived from the colour excess of background stars using the NICER method \citep{Nicer}, are shown in Fig. \ref{fig:vertailu}. The maps are at 40$\arcsec$ resolution. On the right is a masked map of the J-band surface brightness showing the area of the filament with $\tau$(J) > 0.2. The images cover an area of $0.5^{\circ} \times 0.5^{\circ}$.

The observed intensities for the J, H, K, and 3.6 $\mu$m bands were acquired by taking average values over two small areas of the filament, one from the densest part of the filament and one from a part with less optical depth. The locations are marked in the middle panel of figure \ref{fig:vertailu} with white and blue markers, respectively. The obtained intensities are listed in table \ref{tab:observations}.

The above method was also used to acquire values for the observed optical depths. We utilized two optical depth maps derived by \citet{Malinen}, $\tau_{250}$ derived from the Herschel colour temperature and intensity maps with spectral index $\gamma = 1.8$. The second map, $\tau_{J}$, was derived from the near-infrared colour excess following the $NICER$ method described by \citet{Nicer}. The obtained optical depths were $\tau_{\rm J} = 6.10$ and 2.05.

\begin{table}
\caption{The observed surface brightnesses and background intensities used in our computations. The values for the J, H, and K band backgrounds were adopted from \citet{Malinen}}
\label{tab:observations}
\centering
\begin{tabular}{c c c c}
\hline\hline
& $I(\rm \tau_{\rm J} = 6)$ & $I(\rm \tau_{\rm J} = 2)$ & $I_{\rm BG}$ \\
Band & MJy/sr & MJy/sr & MJy/sr \\
\hline
J    & 0.072 & 0.062 & 0.122 \\
H    & 0.152 & 0.121 & 0.100 \\
K    & 0.117 & 0.069 & 0.088 \\
3.6 $\mu$m  & 0.053 & 0.027 & 0.104 \\
\hline
\end{tabular}
\end{table}

\section{The shape of the ISRF}

\begin{figure*}
\sidecaption
\includegraphics[width=12cm]{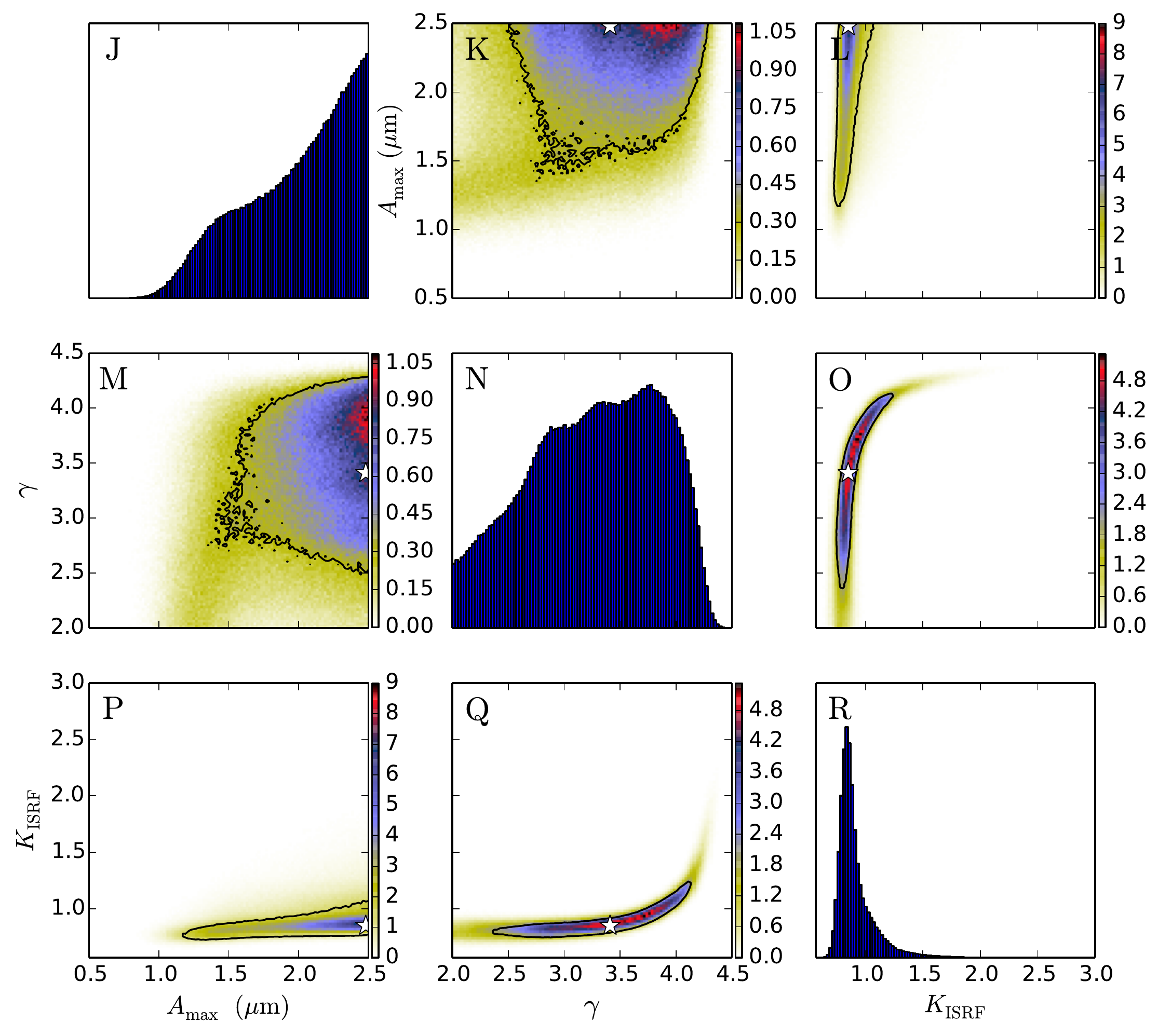} 
\caption{Marginalised probability distributions of dust parameters for the case $\tau_{\rm J}$. For the radiative transfer computations, a one-dimensional cloud model and a three-dimensional radiation field is used. The white star indicates the projected position of the $\chi^2$ minimum. and the black contour shows the 1 $\sigma$ of the projection. The colour scale shows the normalised probability.}
\label{fig:center_3D_isrf_10}
\end{figure*}

In order to distinguish between the effects caused by the ISRF model and the cloud model, we compute a simulation with a one-dimensional cloud model combined with a three-dimensional anisotropic ISRF model and assuming $\sigma_{\rm NIR} = 10\%$. The resulting marginalised probability distributions are shown in Fig. \ref{fig:center_3D_isrf_10}. 

\section{Additional MCMC simulations}

\begin{figure*}
\sidecaption
\includegraphics[width=12cm]{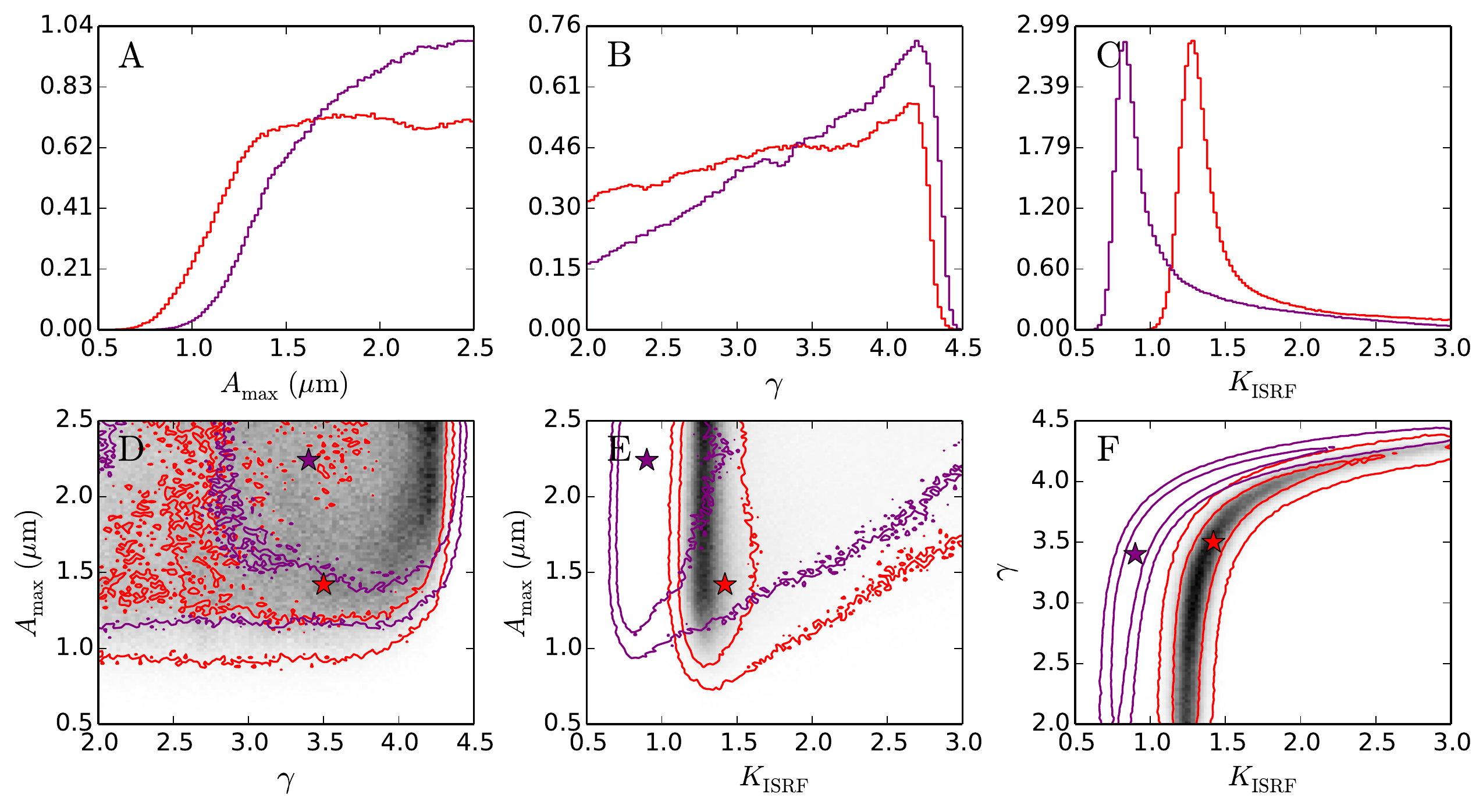} 
\caption{Marginalised probability distributions of dust parameters for the TMC-1N $\tau_J =  6$ position. The grain albedo in the radiative transfer calculations is changed by -10\% (red lines) or +10\% (purple lines), compared to Fig. \ref{fig:centre_control_10_20}. For the surface brightness data $\sigma_{\rm NIR} = 20\%$ is assumed. The stars indicate the projected positions of the $\chi^2$ minima. The grey-scale map corresponds to the red lines and the contours show the 1 and 2 $\sigma$.}
\label{fig:centre_albedo_m_p_20}
\end{figure*}

\begin{figure*}
\sidecaption
\includegraphics[width=12cm]{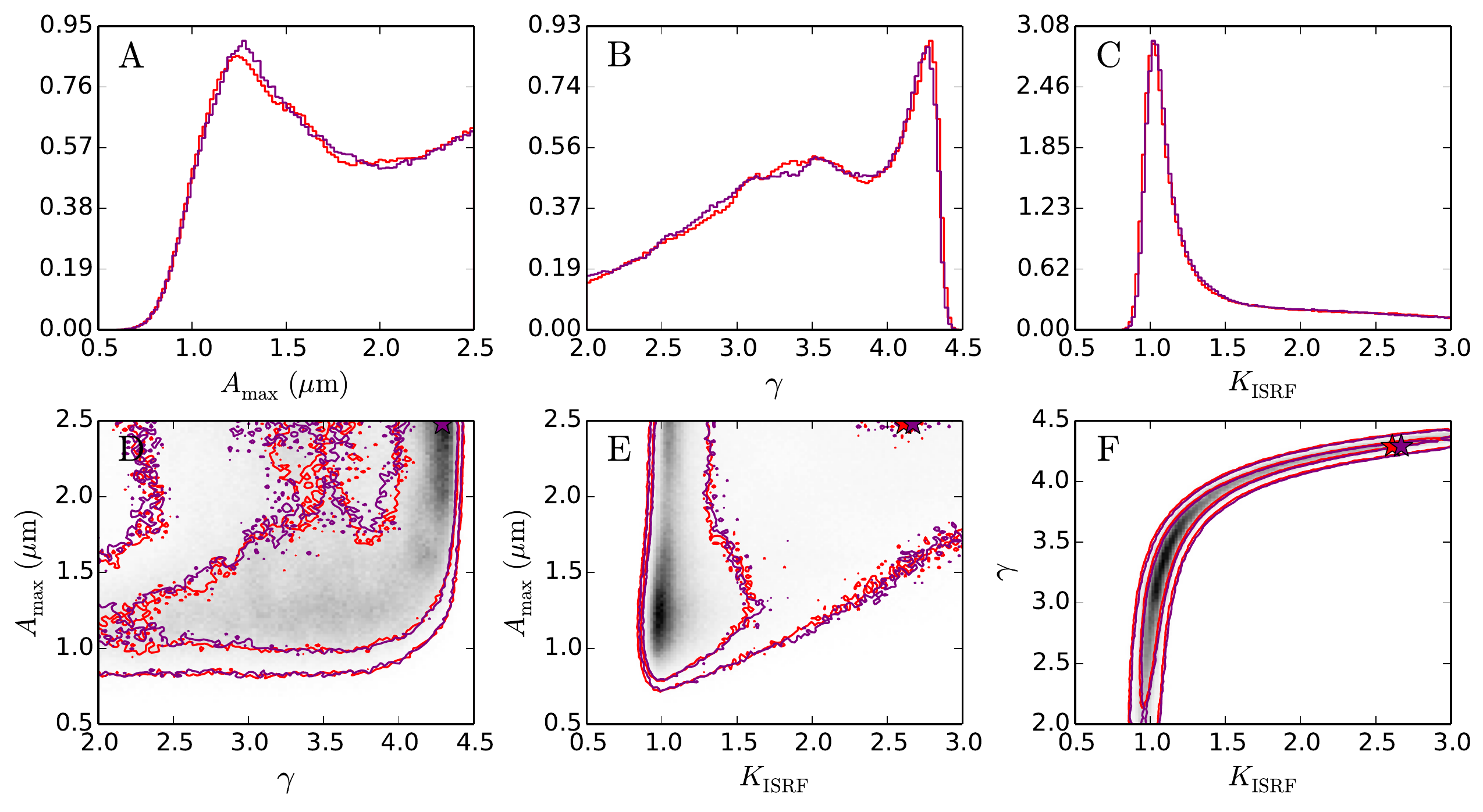} 
\caption{Marginalised probability distributions of dust parameters for the TMC-1N $\tau_J =  6$ position. The asymmetry parameter of the grains in the radiative transfer calculations is changed by -10\% (red lines) or +10\% (purple lines), compared to Fig. \ref{fig:centre_control_10_20}. The stars indicate the projected positions of the $\chi^2$ minima. The grey-scale map corresponds to the red lines and the contours show the 1 and 2 $\sigma$.}
\label{fig:centre_g_m_p_10}
\end{figure*}

Shown in Figs. \ref{fig:centre_albedo_m_p_20} and \ref{fig:centre_g_m_p_10}, are the marginalised probability distributions for TMC-1N $\tau_J =  6$ position where, in the radiative transfer computations, the albedo or the asymmetry parameter of the dust grains is increased, red lines, or decreased, purple lines, by $10\%$. For Fig. \ref{fig:centre_albedo_m_p_20} $\sigma_{\rm NIR} = 20\%$ is assumed and for Fig. \ref{fig:centre_g_m_p_10} $\sigma_{\rm NIR} = 10\%$ is assumed. For both computations $r_{\rm Si}$=0.5 is used.

\begin{figure*}
\sidecaption
\includegraphics[width=12cm]{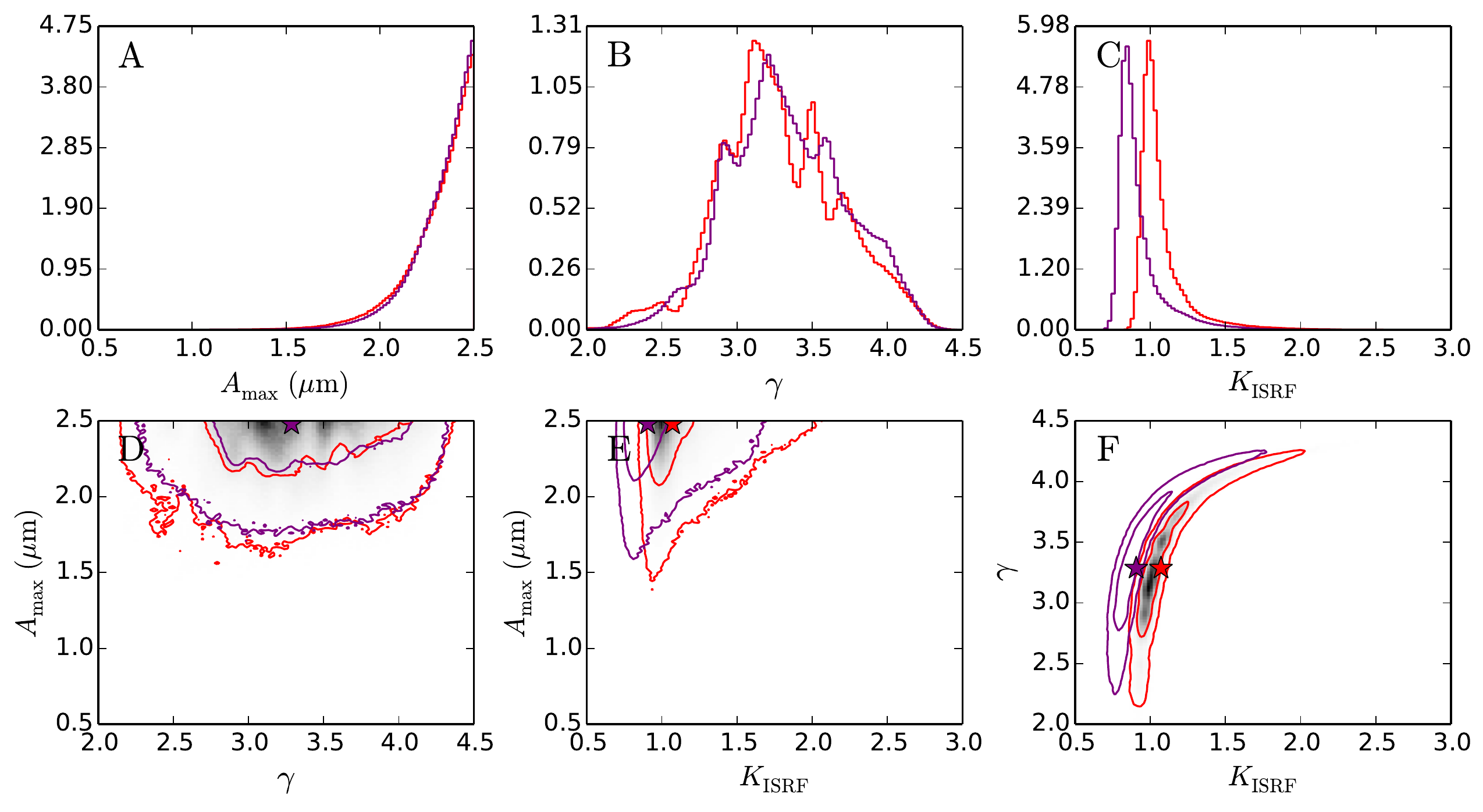} 
\caption{Marginalised probability distributions of dust parameters for the TMC-1N $\tau_J = 2$ position. The grain albedo in the radiative transfer calculations is changed by -10\% (red lines) or the asymmetry parameter has been changed by +10\% (purple lines), compared to Fig. \ref{fig:centre_control_10_20}. The stars indicate the projected positions of the $\chi^2$ minima. The grey-scale map corresponds to the red lines and the contours show the 1 and 2 $\sigma$.}
\label{fig:near_albedo_g_10}
\end{figure*}

Shown in Fig. \ref{fig:near_albedo_g_10} are the marginalised probability distributions for TMC-1N $\tau_J = 2$ position with the asymmetry parameter of the dust grains is changed by +$10\%$ (red lines) or decreased by -$10\%$ (purple lines) in the radiative transfer calculations. For both computations we use $r_{\rm Si}$=0.5, and $\sigma_{\rm NIR} = 10\%$ is assumed.

\end{appendix}
\end{document}